\documentclass[11pt ]{article}

\usepackage{graphicx}
\usepackage{amsmath, amsthm, amssymb}
\usepackage{subfigure}
\usepackage{comment}
\usepackage{bm}
\usepackage{epsfig,color}
\usepackage[sans]{dsfont}

\newcommand{\tr}{\text{tr}}

\newcommand{\be}{\begin{equation}}
\newcommand{\ee}{\end{equation}}
\newcommand{\bea}{\begin{eqnarray}}
\newcommand{\eea}{\end{eqnarray}}
\newcommand{\bes}{\begin{equation*}}
\newcommand{\ees}{\end{equation*}}
\newcommand{\beas}{\begin{eqnarray*}}
\newcommand{\eeas}{\end{eqnarray*}}

\newcommand{\x}{\mathrm{x}}

\renewcommand{\H}{\mathcal{H}}

\def\B{\mathcal{B}}

\def\x{\mathrm{x}}
\def\y{\mathrm{y}}
\def\g{\mathrm{guess}}

\def\K{\widetilde{K}}
\def\v{\vec{v}}

\def\P{\mathrm{P}}
\def\tr{\mathrm{tr}}

\newtheorem{thm}{Theorem}[section]
\newtheorem*{thm*}{Theorem}

\newtheorem{lem}[thm]{Lemma}
\newtheorem*{lem*}{Lemma}
\newtheorem{prop}[thm]{Proposition}
\newtheorem{defn}[thm]{Definition}
\newtheorem{rem}[thm]{Remark}

\newtheorem*{lipschitzLem*}{Lemma \ref{lipschitz}}
\newtheorem*{lipschitzCubeLem*}{Lemma \ref{lipschitzCube}}
\newtheorem*{pgmNearlyOptimalThm*}{Theorem \ref{pgmNearlyOptimal}}

\begin{document}

\title{Structure of minimum-error quantum state discrimination}

\date{\today} 

\author{Joonwoo Bae \footnote{E-mail: bae.joonwoo@gmail.com} \\ \\[0.5em]
{\it\small  Center for Quantum Technologies, National University of Singapore, Singapore 117543,
}\\
{\it\small  ICFO--Institut de Ci\`encies Fot\`oniques, 08860 Castelldefels (Barcelona), Spain.}}

\maketitle

\begin{abstract}
Distinguishing different quantum states is a fundamental task having practical applications for information processing. Despite the efforts devoted so far, however, strategies for optimal discrimination are known only for specific examples. We here consider the problem of minimum-error quantum state discrimination where the average error is attempted to be minimized. We show the general structure of minimum-error state discrimination as well as useful properties to derive analytic solutions. Based on the general structure, we present a geometric formulation of the problem, which can be applied to cases where quantum state geometry is clear. We also introduce equivalent classes of sets of quantum states in terms of minimum-error discrimination: sets of quantum states in an equivalence class share the same guessing probability.  In particular, for qubit states where the state geometry is found with the Bloch sphere, we illustrate that for an arbitrary set of qubit states, the minimum-error state discrimination with equal prior probabilities can be analytically solved, that is, optimal measurement and the guessing probability are explicitly obtained. 
\end{abstract}



\section{Introduction}

In the early days when quantum systems are applied to communication of distant parties, in particular photonic systems which define physical limits on sources of optical communication, there have been pioneering works that investigate how quantum systems can be exploited to information processing and incorporated to frameworks of classical information theory \cite{ref:holevo} \cite{ref:yuen} \cite{ref:hel}. One of the most fundamental tasks for such applications is to formalize how classical messages can be retrieved from quantum states. For various purposes, different designs of measurements have been then introduced and studied, for instance, measurement settings are optimized for accessible information \cite{ref:sas}, unambiguous discrimination \cite{ref:iva} \cite{ref:die} \cite{ref:per} \cite{ref:jae} \cite{ref:ban}, maximizing confidence to guess about certain states \cite{ref:cro}, or minimizing errors on average when making guesses about given states \cite{ref:holevo} \cite{ref:yuen} \cite{ref:hel}, etc. Recently, there have also been approaches that, with pre-determined rate of inconclusive results, errors in discrimination are minimized \cite{ref:bagan} \cite{ref:herzog}. All these are fundamental and practical in quantum information theory and its applications, as well as useful tools to investigate foundational aspects of quantum theory, see also reviews in Refs. \cite{ref:rev1} \cite{ref:rev2} \cite{ref:rev3} \cite{ref:rev4} \cite{ref:rev5}.
 
Among different strategies in quantum state discrimination, here we are interested in the minimum-error (ME) discrimination. The ME discrimination optimizes measurement so that, once a quantum state is given from a set of known quantum states, one aims to make the correct guess about the state with a minimal error on average. This in fact has a number of applications in quantum information tasks, and is also of a fundamental interest since it shows the ultimate limit in the identification of, among a specified set of states, a given state. 

Desired is then to have, for arbitrarily given sets of quantum states, a general method of finding optimal measurement, that is, measurement that achieves minimal errors on average in the state discrimination. Despite the efforts devoted so far, however, analytic solutions are known only for restricted classes of quantum states, e.g. cases when certain symmetries are contained. The lack of solutions in most cases is partly because the ME discrimination itself has been largely understood as an optimization problem that is generally hard to have analytic solutions. Otherwise, little is known as an approach to the ME discrimination. Needless to speak about the importance in its own right, the lack of a general method for state discrimination has been obviously and also potentially a significant obstacle preventing further investigation in both quantum information theory and quantum foundation. 

To be precise about known results in the ME discrimination, no general method has been known as an analytic way of solving the ME state discrimination. The known general method is a numerical approach via semidefinite programming, which efficiently returns the solution in a polynomial time \cite{ref:jez} \cite{ref:eldarsdp2} \cite{ref:eldar-sdp} \cite{ref:sdpbook}. For cases in which the ME discrimination is known in an analytic form, two-state discrimination is the only case where no symmetry is assumed among given quantum states. The result was shown in 1976 and called as the Helstrom bound \cite{ref:hel}. Apart from this, if no symmetry exists among given states, no analytic solution is known, for instance, in the next simplest case of arbitrary three states, or even simpler case of three qubit states. Otherwise, the ME discrimination is known for cases when given quantum states have certain symmetries, such as the geometrically uniform structure, see Refs. \cite{ref:hel} \cite{ref:sas} \cite{ref:eldarsdp2} \cite{ref:mirror3}  \cite{ref:eldar} \cite{ref:mirror0} \cite{ref:mirror4} \cite{ref:mirror1} \cite{ref:mirror2}.

Note also that, although general solutions are lacked, the necessary and sufficient conditions that characterize optimal discrimination have been obtained in the very beginning when the problem was introduced \cite{ref:holevo} \cite{ref:yuen}. That is, parameters giving optimal discrimination satisfy the conditions, and conversely, any parameters fulfilling the conditions can construct optimal discrimination. They have been used to run a numerical algorithm \cite{ref:jez}.

The purpose of the present work is to show the general structure existing in the ME discrimination, namely, the relations among optimal measurement, the guessing probability and other useful properties to analytically find optimal measurement and the guessing probability. The main idea is to view the ME discrimination from various angles and different approaches, such as relations of fundamental principles, convex optimization frameworks, or their generalization called the complementarity problem, and different formalisms accordingly. We show that all these can be equivalently summarized into the so-called complementarity problem in the context of convex optimization, see e.g. Ref. \cite{ref:com}. The approach refers to a direct analysis on optimality conditions: in fact, in this way, more parameters are included than originally given problems, however, the advantage is that the generic structure of given optimization problems is fully exploited. Applying the structure and the method of the complementarity problem, we provide a geometric formulation of solving the ME discrimination. Thus, once the geometry is clear from the context of given quantum states, the formulation can be exploited and one can straightforwardly find the solution such as optimal measurement and the guessing probability. An an example where the state geometry is also clear, we illustrate that the ME discrimination is completely solved for an arbitrary set of qubit states given with equal prior probabilities. All these provide alternative methods of solving the ME discrimination in an analytic way, apart from the numerical optimization method.

The rest of the paper is structured as follows. In Sec. \ref{sec:optimal}, we begin with the problem definition of and introduction to the ME discrimination. We also provide an operational interpretation of the guessing probability. We obtain optimality conditions for the ME discrimination from different approaches, and finally put them equivalently into the framework of complementarity problem. From the optimality conditions, we show the general structure of optimal parameters in the ME discrimination, such as optimal measurement and the guessing probability. In this way, we provide a geometric formulation the ME discrimination. In Sec. \ref{sec:general}, based on the general structure, we present general properties as follows. We construct equivalence classes of sets of quantum states in terms of the ME discrimination, such that for the sets in the same class the ME discrimination is characterized in an equivalent way. We also show, conversely, how a set of states can be generated in an equivalent class. Then, for the generated states, the ME discrimination is already characterized by the equivalence class. We finally present various and equivalent forms of the guessing probability according to different approaches that the general structure and the optimality conditions are derived, and provide their operational meanings. In Sec. \ref{sec:qubit}, we apply all these to qubit states in which quantum state geometry is clearly found with the Bloch sphere. We show how the geometric formulation can be applied and illustrate that for all problems of the ME discrimination with equal prior probabilities can be analytically solved by using the state geometry. In Sec. \ref{sec:con}, we conclude the results. 


\section{Minimum-error discrimination and optimality conditions}
\label{sec:optimal}

In this section, let us introduce the problem of ME discrimination of quantum states and fix notations. We then analyze the optimality conditions that completely characterize optimal parameters such as optimal measurement and the guessing probability. We review optimality conditions derived from different approaches and show that they are equivalent. 


\subsection{Problem definition and probability-theoretic preliminaries}
\label{sec:pre}

Suppose that there is a device which generates different quantum states. The device has $N$ buttons, and pressing one of them, say $\x$, corresponds to an instance that a state $\rho_{\x}$ is generated. Denoted by $q_{\x}$ as the probability that the button $\x$ is pressed, {\it a priori} probability that $\rho_{\x}$ is produced from the device is defined. We summarize the state generation by $\{q_{\x}, \rho_{\x} \}_{\x=1}^{N}$. Once generated, a state is then sent to a measurement device, which is prepared to make a best guess about which button has been pressed in the preparation. This corresponds to the task of distinguishing $N$ quantum states via measurement devices.

Let $P(\x|\y)$ denote the probability that when state $\rho_{\y}$ has been prepared, one concludes from measurement outcomes that state $\rho_{\x}$ is given. Or, equivalently as the conclusion follows from measurement outcomes, it is the probability that an output port $\x$ is clicked when a button $\y$ is pressed in the preparation. Then, in the ME discrimination, we are interested in minimizing the probability of making errors on average, or equivalently to have maximal probability to make correct guesses about input states, called \emph{the guessing probability},  $\P_{\g} = \max_M \sum_{\x} q_{\x} P(\x | \x)$ where the maximization runs over measurement settings $M$. 

Measurement on quantum systems is described by Positive-Operator-Valued-Measures (POVMs): a set of positive operators $\{ M_{\x} \geq 0\}_{\x = 1}^{N}$ that fulfills, $\sum_{\x = 1}^{N} M_{\x} = I$. For given states $\{q_{\x} , \rho_{\x} \}_{\x=1}^{N}$, let $\{M_{\x} \}_{\x=1}^{N}$ denote POVMs for quantum state discrimination. When state $\rho_{\y}$ is actually given, the probability that a detection event happens in measurement $M_{\x}$ is given by $\P(\x|\y) =\tr[\rho_{\y}M_{\x}]$. The guessing probability can therefore be written as
\bea 
\P_{\mathrm{guess}} = \max_M \sum_{\x=1}^{N}  q_{\x} P(\x | \x) =  \max_{\{ M_{\x} \}_{\x=1}^{N}} \sum_{\x = 1 }^{N} q_{\x} \tr[M_{\x}\rho_{\x}], \label{eq:guessprob} 
\eea 
with the maximization over all POVMs. From postulates of quantum theory, non-orthogonal quantum states cannot be perfect discriminated between, i.e. $P_{\g}<1$. 

The ME discrimination described in the above can be rephrased in a probability-theoretic way as follows. The preparation applies random variable $X$, which corresponds to a button pressed, and the measurement shows the other random variable $Y$ for the guessing with some probabilities. Distinguishability of different probabilistic  systems can be in general measured by variational distance of probabilities. Then, the distance from the uniform distribution is particularly useful to describe and interpret the guessing probability. The distance of a probability distribution of random variable $P_{X}$ from the uniform distribution, which is $1/N$ in this case, is denoted by $d(P_X)$ or simply $d(X)$ and defined as
\bea
d(X) = \frac{1}{2} \sum_{\x=1}^{N} \big | P_{X}( \x )  - \frac{1}{N}  \big|.  \label{eq:disuni}
\eea
Let $d(X|Y)$ denote the distance of random variable $X$ given $Y$ from the uniform.


\begin{lem}
\label{lem:2-1}
\cite{ref:maurer}  The guessing probability about $X$ given $Y$ can be expressed by
\bea
P_{\g} := P_{\g}(X|Y) = \frac{1}{N} + d(X |Y).  \label{eq:lem2-1}
\eea
\end{lem}

\emph{Proof.} Note that the guessing probability is, $P_{\g} = \sum_{\y=1}^{N} P_{Y}(\y) P_{X|Y} (\y|\y)$. It is then straightforward to write  the distance, as follows,
\bea
d(X|Y) & = & \sum_{\y=1} P_{Y}(\y) d(X| Y = \y) \nonumber\\
& = &  \frac{1}{2} \sum_{\y=1}^{N} P_{Y}(\y) \sum_{\x=1}^{N}  \big|  P_{X|Y} (\x | \y) -\frac{1}{N} \big|. \nonumber
\eea
As a probability of making correct guesses is not smaller than the random guess, we can assume that $P_{X|Y}(\y|\y) \geq 1/N$ for all $\y=1,\cdots,N$ and $P_{X|Y}(\x|\y) \leq 1/N$ for $\x \neq \y$, and this does not lose any generality. Thus, it follows that, when $Y=\y$,
\bea
\sum_{\x=1}^{N}  \big|  P_{X|Y} (\x | \y) -\frac{1}{N} \big| & = & P_{X|Y} (\y|\y) -\frac{1}{N} - \sum_{\x \neq\y} (P_{X|Y}(\x | \y) -\frac{1}{N}) \nonumber\\
& = & 2 P_{X|Y} (\y | \y) -\frac{2}{N}\nonumber
\eea
From these two relations, it holds that,
\bea 
d(X|Y) = P_{\g} - \sum_{\y=1}^{N} P_{Y}(\y) \frac{1}{N}  = P_{\g}  -  \frac{1}{N}  \nonumber
\eea
and thus Eq. (\ref{eq:lem2-1}) is shown. $\Box$\\

This shows that the guessing probability about random variable $X$ given $Y$ corresponds to the distance of the probability $P_{X|Y}$ from the uniform distribution. As only input and output random variables are taken into account, the expression in Eq. (\ref{eq:lem2-1}) generally works for any physical systems, either quantum or classical, employed from preparation to measurement. Then, the distance $d(X|Y)$ depends on physical systems employed between preparation and measurement. Once quantum systems are applied, the distance $d(X|Y)$ must be expressed in terms of properties or relations of given quantum states. The expression for quantum systems is to be shown in Lemma \ref{lem:qguess}.


\subsection{Optimality conditions}
\label{subsec:opt} 

As it is shown in Eq. (\ref{eq:guessprob}), optimal discrimination is obtained with POVM elements that maximize the probability of making correct guesses. There have been different approaches to characterize optimal measurement. Here, we review known results and show that they are equivalent one another.

\subsubsection*{i) Optimality conditions from analytic derivation} 

The necessary and sufficient conditions that POVMs must satisfy to fulfill optimal discrimination have been obtained from the very beginning when the problem was introduced in Refs. \cite{ref:holevo} \cite{ref:yuen}. For given states $\{q_{\x},\rho_{\x} \}_{\x=1}^{N}$ to discriminate among, POVM elements $\{ M_{\x}\}_{\x = 1}^{N}$ achieve the guessing probability if and only if they satisfy
\bea
M_{\x} (q_{\x} \rho_{\x} - q_{\y} \rho_{\y}) M_{\y} & = & 0, ~~\forall \x, \y=1,\cdots, N \label{eq:opt1}\\
\sum_{\x =1 }^{N} q_{\x} \rho_{\x } M_{\x } - q_{\y} \rho_{\y} & \geq & 0, ~~\forall  \y=1,\cdots, N.   \label{eq:opt2}
\eea 
That is, for a state discrimination problem, if one finds a set of POVMs that satisfy two conditions above, then optimal discrimination is immediately obtained. Note also that for a given problem of state discrimination, optimal measurement is generally not unique. 

\subsubsection*{ii) Optimality condition from convex optimization}

The optimality conditions can also be derived from a different context, the formalism of convex optimization \cite{ref:sdpbook}. The optimization problem in Eq. (\ref{eq:guessprob}) with the convex and affine constraints on POVMs can be written as,
\bea
\max && \sum_{\x =1}^{N} q_{\x} \tr[M_{\x} \rho_{\x}] \nonumber \\
\mathrm{subject} ~\mathrm{to} && \sum_{\x =1 }^{N} M_{\x} = I,~ \mathrm{and}~ M_{\x} \geq 0,~\forall~ \x =1,\cdots, N. \label{eq:primal}
\eea
For convenience, we call Eq. (\ref{eq:primal}) the primal problem. Note that the primal problem in the above is feasible. 

To derive its dual problem, let us first construct the Lagrangian $\mathcal{L}$ as follows,
\bea
\mathcal{L}(\{M_{\x} \}_{\x=1}^{N}, \{\sigma_{\x} \}_{\x=1}^{N},K) &=& \sum_{\x=1}^{N} q_{\x} \tr[M_{\x} \rho_{\x}] + \sum_{\x=1}^{N} \tr[\sigma_{\x} M_{\x}] + \tr[K (I - \sum_{\x=1}^{N} M_{x})], \nonumber \\
& = & \sum_{\x=1}^{N} \tr[M_{\x} (q_{\x} \rho_{\x} +\sigma_{\x} -K ) ] + \tr{K} \nonumber
\eea
where $\{\sigma_{\x} \}_{\x=1}^{N}$ are positive-semidefinite Hermitian operators and $K$ is a Hermitian operator. Note that $\{\sigma_{\x} \}_{\x=1}^{N}$ and $K$ are called dual parameters. To derive the dual problem, we first have to maximize the Lagrangian over primal parameter  $\{ M_{\x} \}_{\x=1}^{N}$.  If it holds that $q_{\x}\rho_{\x} +\sigma_{\x} -K > 0 $ for some $\x$, then the primal problem becomes not feasible since the maximization may go to $+\infty$. Thus, we have $q_{\x} \rho_{\x} +\sigma_{\x} -K \leq 0$ for all $\x$, where the equality holds if and only if the Langrangian is maximized. Combined together with the other constraint that $\sigma_{\x}\geq 0$ for all $\x$, the constraint can be finally rewritten as, $K\geq q_{\x} \rho_{\x}$ for all $\x$. This also shows that the operator $K$ must be positive. Note that since Lagrangian is maximized, we now have
\bea
\max_{\{ M_{\x} \}_{\x=1}^{N}} \sum_{\x =1}^{N} q_{\x} \tr[M_{\x} \rho_{\x}]  \leq \max_{\{ M_{\x} \}_{\x=1}^{N}}  \mathcal{L}(\{M_{\x} \}_{\x=1}^{N}, \{\sigma_{\x} \}_{\x=1}^{N},K) \vert_{q_{\x} \rho_{\x} +\sigma_{\x} -K \leq 0}= \tr K, \nonumber
\eea
which is called the weak duality.

The problem dual to Eq. (\ref{eq:primal}), called the dual problem, is therefore obtained as follows,
\bea 
\min && \tr[K] \nonumber \\
\mathrm{subject} ~\mathrm{to} && K \geq q_{\x} \rho_{\x},~ \forall~\x=1,\cdots, N. \label{eq:dual}
\eea
Putting optimal discrimination problems into this convex optimization framework, the solution i.e. the guessing probability is returned efficiently in a polynomial time. In general, solutions from primal and dual problems in the convex optimization are not necessarily the same. It generally holds that either of solutions is larger than or equal to the other, from the property called weak duality as it is shown in the above. In this problem of state discrimination, it turns out that solutions from both problems coincide each other, which follows from so-called constraint quantification. This property is called the strong duality. Thus, we have 
\bea
\P_{\g} = \max_{ \{  M_{\x} \}_{\x=1}^{N} } \sum_{\x =1}^{N} q_{\x} \tr[M_{\x} \rho_{\x}]  = \min_{K\geq q_{\x} \rho_{\x} } \tr[K]. \nonumber
\eea
That is, the guessing probability is obtained by solving either primal or dual problem.

Apart from solving either the primal or the dual problem, a third approach in  the convex optimization, which we are mainly to consider here, is the so-called {\it complementarity problem} that directly analyzes constraints characterizing optimal parameters in both primal and dual problems and the finds all of optimal parameters \cite{ref:bae}. As both of primal and dual parameters from both convex constraints are taken into account, the approach itself is not considered to be more efficient in numerics than primal or dual problems. Its advantage, however, lies at the fact that general structure existing in an optimization problem are fully exploited. 

Those constraints which characterize optimal solutions are called optimality conditions, and can be written as the so-called Karush-Kuhn-Tucker (KKT) conditions. In general, KKT conditions are only necessary since solutions of primal and dual problems can be unequal. As we have mentioned, from the constraint qualification the strong duality holds, and consequently KKT conditions in this case are also sufficient. The list of KKT conditions for the discrimination problem is, then, constraints in Eqs. (\ref{eq:primal}) and (\ref{eq:dual}) and two more conditions in the following.

\begin{lem}
\label{lem:opt}
(Optimality conditions) \\
For a set of states $\{q_{\x},\rho_{\x} \}_{\x=1}^{N}$, the optimal ME discrimination is characterized by a symmetry operator denoted by $K^{\star}$ and complementary states $\{ r_{\x}^{\star} , \sigma_{\x}^{\star} \}_{ \x = 1}^{N}$, where $r_{\x}^{\star} \geq 0$, which satisfy the followings, for $\x = 1,\cdots,N$,
\bea
\mathrm{(symmetry}~ \mathrm{operator)}~~~~~~~~~~~ K^{\star} & = & q_{\x} \rho_{\x} + r_{\x}^{\star} {\sigma}_{\x}^{\star}, \label{eq:symop} \\ 
\mathrm{(orthogonality)} ~~~ r_{\x}^{\star} \tr[M_{\x}^{\star}{\sigma}_{\x}^{\star}] &=& 0, \label{eq:orth} 
\eea
where POVM elements $\{ M_{\x}^{\star} \}_{\x=1}^{N}$ in the above are an optimal measurement giving the guessing probability. 

Conversely, for given states $\{q_{\x},\rho_{\x} \}_{\x=1}^{N}$, any set of parameters, $K$, $\{ r_{\x}, \sigma_{\x}\}_{\x=1}^{N}$, and $\{M_{\x} \}_{\x=1}^{N}$ fulfilling KKT conditions, Eqs. (\ref{eq:primal}), (\ref{eq:dual}), (\ref{eq:symop}), and (\ref{eq:orth}), characterize the optimal state discrimination. If parameters satisfy KKT conditions, they are optimal and we write them by $K^{\star}$, $\{ r_{\x}^{\star}, \sigma_{\x}^{\star} \}_{\x=1}^{N}$, and $\{M_{\x}^{\star} \}_{\x=1}^{N}$, throughout. 

Note that, with optimal parameters, the guessing probability is given by $P_{\g} = \tr[K^{\star}]$ from the dual problem in Eq (\ref{eq:dual}). 
\end{lem}

Let us explain how one can obtain two conditions in Eqs. (\ref{eq:symop}) and (\ref{eq:orth}). In the context of convex optimization, the condition of symmetry operator in Eq. (\ref{eq:symop}) follows from the Lagrangian stability: $\nabla_{M_{\x}}\mathcal{L}=0$ for all $\x$. The orthogonality condition in Eq. (\ref{eq:orth}) is obtained from the complementary slackness \cite{ref:sdpbook}. In state discrimination, therefore, those parameters satisfying KKT conditions, Eqs. (\ref{eq:primal}), (\ref{eq:dual}), (\ref{eq:symop}) and (\ref{eq:orth}) are characterized as optimal measurements and complementary states to give guessing probabilities. 

We in particular call $K^{\star}$ as the symmetry operator in the ME discrimination of states $\{ q_{\x} ,\rho_{\x}\}_{\x=1}^{N}$ due to the following reasons. First, the operator is uniquely determined for the ME discrimination of given states, whereas optimal measurement is not unique e.g. Ref. \cite{ref:mochon}. Later, the proof is provided in Lemma \ref{lem:comp}. This means that for given quantum states, the symmetry operator, rather than optimal measurement, characterizes the ME discrimination. Next, due to the uniqueness, the operator preserves the properties concerning about the guessing probability in ME discrimination. Note that the guessing probability is of practical importance when exploiting sets of quantum states in a communication task. Then, as we will explicitly show later in examples, the guessing probability does not generally depend on detailed relations (e.g. angles or distances between states) of given states $\{ q_{\x} ,\rho_{\x}\}_{\x=1}^{N}$ but a single parameter, the operator $K^{\star}$ constructed by given states. 

Let us also be precise about the result that the guessing probability does not depend detailed relations of given quantum states. We point out that the observation from the two-state discrimination \cite{ref:hel}, where the trace-distance defining the relation between the two states is the parameter dictating the guessing probability, cannot generalize to more than three states. In addition, we show that, by examples, it clearly fails to generalize the observation for ME discrimination of more than two states. As it is to be shown in Sec. \ref{sec:qubit}, one of three states can be modified independently, while the guessing probability remains the same. We emphasize that the operator $K^{\star}$ directly dictates and corresponds to the guessing probability, rather than any other parameters in the ME discrimination. We also note that this property is along the conclusion in Ref. \cite{ref:jozsa} that distinguishability is a global property that cannot be reduced to distinguishability of each pair of states. 

In fact, two distinct sets of quantum states can have the same symmetry operator. Then, since the symmetry operator gives the complete characterization of the ME discrimination such as the guessing probability and complementarity states, the ME discrimination for the two sets is analyzed in terms of the same symmetry operator. This motivates to construct equivalence classes of sets of quantum states via the symmetry operator in the ME discrimination, and shows a general structure in the ME discrimination, see Sec. \ref{sec:general}. 

In the below, we show that optimality conditions in Eqs. (\ref{eq:symop}) and (\ref{eq:orth}) are equivalent to those in Eqs. (\ref{eq:opt1}) and (\ref{eq:opt2}).   \\

\begin{rem}
KKT conditions in Eqs. (\ref{eq:symop}) and (\ref{eq:orth}) are equivalent to the optimality conditions shown in Eqs. (\ref{eq:opt1}) and (\ref{eq:opt2}).
\end{rem}

\emph{Proof.} To prove the equivalence, it suffices to show that KKT conditions imply optimality conditions in Eqs. (\ref{eq:opt1}) and (\ref{eq:opt2}). In the following, we derive Eqs. (\ref{eq:opt1}) and (\ref{eq:opt2}) from KKT conditions. 

First, since the symmetry operator in Eq. (\ref{eq:symop}) gives the guessing probability, we have
\bea 
P_{\g} =  \tr[K^{\star}] =   \tr[\sum_{\x=1}^{N} q_{\x} \rho_{\x} M_{\x}^{\star}]. \label{eq:lem2-2}
\eea
It follows that $K^{\star} =  \sum_{\x=1}^{N} q_{\x} \rho_{x} M_{\x}^{\star}$ with optimal POVMs fulfilling Eq. (\ref{eq:lem2-2}). From Eq. (\ref{eq:symop}), noting that $r_{\x}^{\star} \sigma_{\x}^{\star} \geq 0$, we have $K^{\star} - q_{\x}\rho_{\x} \geq 0$ for all $\x=1,\cdots,N$. This already proves the condition in Eq. (\ref{eq:opt2}).

Next, from the symmetry operator in Eq. (\ref{eq:symop}), the equality in Eq. (\ref{eq:lem2-2}) also implies the following:
\bea
&& \tr[K^{\star}] = \tr[\sum_{\x=1}^{N} q_{\x} \rho_{\x} M_{\x}^{\star}  ] = \tr[\sum_{\x=1}^{N} (K^{\star} - r_{\x}^{\star} \sigma_{x}^{\star} ) M_{\x}^{\star}] = \tr[K^{\star}] -   \sum_{\x=1}^{N} \tr [ r_{\x }^{\star} \sigma_{\x}^{\star} M_{\x}^{\star} ] \nonumber \\
&\Rightarrow &  \sum_{\x=1}^{N} \tr [ r_{\x }^{\star} \sigma_{\x}^{\star} M_{\x}^{\star} ] = 0.\nonumber
\eea
Since $\tr[ r_{\x}^{\star} \sigma_{\x}^{\star} M_{\x}^{\star} ]\geq0 $ for each $\x$, we conclude that $\tr [ M_{\x}^{\star} r_{\x}^{\star} \sigma_{\x}^{\star} ] = 0$ for all $\x$. Moreover, since $\sigma_{\x}^{\star} \geq 0$, $M_{\x}^{\star} \geq 0$, and $r_{\x}^{\star} \geq 0$, we have $ r_{\x}^{\star}    \sigma_{\x}^{\star} M_{\x}^{\star} =0$ for all $\x$. We apply this identity to the following, together with the relation $q_{\x}\rho_{\x} = K^{\star} - r_{\x}^{\star} \sigma_{\x}^{\star}$ in Eq. (\ref{eq:symop}): 
\bea 
M_{\x}^{\star} (q_{\x} \rho_{\x}  -  q_{\y} \rho_{\y}) M_{\y}^{\star} & = &  M_{\x}^{\star} ( (K^{\star} - r_{\x}^{\star} \sigma_{\x}^{\star} ) -   ( K^{\star} - r_{\y}^{\star} \sigma_{\y}^{\star} )   ) M_{\y}^{\star} \nonumber \\
& = &M_{\x}^{\star} ( r_{\y}^{\star} \sigma_{\y}^{\star} M_{\y}^{\star} ) -  ( M_{\x}^{\star} r_{\x}^{\star} \sigma_{\x}^{\star}) M_{\y}^{\star} =0. \nonumber
\eea
Thus, it is shown that the KKT conditions imply the optimality condition in Eq. (\ref{eq:opt1}). $\Box$\\

\subsubsection*{iii) Optimality conditions from fundamental principles} 
 \label{subsec:fun}
 
In Ref. \cite{ref:bae11}, it is shown that the no-signaling principle can generally determine the ME quantum state discrimination. This is shown by proving that if measurement devices are non-signalling, the guessing probability in the state discrimination cannot be larger than what is characterized within quantum theory. Here, we briefly sketch the main framework of the derivation and show that the optimality conditions in Eqs. (\ref{eq:symop}) and (\ref{eq:orth}) can be alternatively derived from the fundamental principles. 

Two operational tasks in quantum theory are exploited. One is the \emph{ensemble steering}, which states that there always exists a bipartite quantum state such that sharing the state, one party can prepare any ensemble decomposition of the other party's state. This was firstly asserted by Schr\"odinger \cite{ref:schr} and later formalized as Gisin-Hughston-Jozsa-Wootters (GHJW) theorem \cite{ref:ghjw1} \cite{ref:ghjw2}. The other is the no-signaling principle that information cannot be transmitted faster than the speed of light. Then, the result in Ref. \cite{ref:bae11} is that optimal quantum state discrimination is a consequence that two correlations are compatible, i) one from quantum correlations, called ensemble steering, that ensemble decompositions of quantum states can be steered by a party at a distance, and ii) the other, the no-signaling condition on probability distributions, that probability distributions of input and output random variables at a location cannot be exploited to instantaneous communication. From two fundamental principles, optimality conditions in Eqs. (\ref{eq:symop}) and (\ref{eq:orth}) can be reproduced as follows. 

Suppose that two parties share entangled states, and one party prepares quantum states (to be discriminated among) to the other far in distance by using ensemble steering. For states $\{q_{\x}, \rho_{\x} \}_{\x=1}^{N}$ for the ME discrimination, the other parties are with the following identical ensemble in different decompositions, 
\bea 
\rho = p_{\x} \rho_{\x} + (1-p_{\x}) \sigma_{\x},~\mathrm{for}~\x = 1,\cdots,N \label{eq:ens}
\eea
for some states $\{ \sigma_{\x}\}_{\x=1}^{N}$ with $q_{\x} = p_{\x} / \sum_{\y = 1}^{N} p_{\y}$. Notice that an identical ensemble $\rho$ has different decompositions according to indices $\x=,1\cdots,N$. Existence of states $\{ \sigma_{\x}\}_{\x=1}^{N}$ that compose the identical ensemble in the above together with a given set $\{ \rho_{\x}\}_{\x=1}^{N}$ follows from the GHJW theorem in Refs. \cite{ref:ghjw1} \cite{ref:ghjw2}. Note that $\{ q_{\x}\}_{\x=1}^{N}$ are prior probabilities and thus $\sum_{\x=1}^{N} q_{\x} =1$, and that $\{ p_{\x}\}_{\x=1}^{N}$ are \emph{probabilities to steer} quantum states $\{ \rho_{\x}\}_{\x=1}^{N}$ in the ensembles and thus we do not necessarily have $\sum_{\x=1}^{N} p_{\x} = 1$.

If a measurement device is set only to discriminate among states $\{ \rho_{\x}\}_{\x=1}^{N}$, the capability of optimal discrimination of $\{q_{\x}, \rho_{\x} \}_{\x=1}^{N}$ cannot work with an arbitrarily high probability. This is because, if the ME discrimination works arbitrarily well, it would violate the no-signaling principle. For instance, suppose that the ensemble decomposition is concluded by the ME discrimination of states $\{\rho_{x} \}_{\x=1}^{N}$: that is, concluding that $\rho_{\x}$ is found in the discrimination, one guesses that the ensemble corresponds to $p_{\x}\rho_{\x} + (1- p_{\x})\sigma_{\x} $. While no information is announced about the preparation of ensemble decompositions among $\x=1,\cdots,N$, the no-signaling condition is fulfilled, and therefore the ME discrimination of $\{\rho_{\x} \}_{\x=1}^{N}$ must not allow one to gain knowledge about the preparation. Given no information about the preparation, it must be a random guess about ensemble decompositions. Thus, state discrimination can be constrained such that the strategy above gives at its best the random guess.

In Ref. \cite{ref:bae11} it is shown that the ME discrimination must satisfy the following condition to fulfill two constraints in the above,
\bea
\sum_{\x=1}^{N} p_{\x} P (\x | \x, \{ \rho_{\x}\}_{\x=1}^{N}   ) \leq 1,  \label{eq:bdnos1}
\eea
where $ P (\x | \x, \{ \rho_{\x}\}_{\x=1}^{N}   )$ is the probability of giving outcome $\x$ when one of states $\{ \rho_{\x}\}_{\x=1}^{N}$ is given i.e. the probability of correctly discriminating among states $\{ \rho_{\x} \}_{\x=1}^{N}$. The equality holds if and only if, for an ensemble prepared in Eq. (\ref{eq:ens}) with some $\x$, a measurement device responds only to quantum states $\rho_{\x} $ but not to states $\sigma_{\x}$, that is, $P(\x | \x,   \{ \sigma_{\x}\}_{\x=1}^{N}  ) = 0$. Note that, with the measurement postulate, this is equivalent to $\tr[M_{\x}\sigma_{\x}]=0$ for some POVM $M_{\x}$, which is indeed the optimality condition in Eq. (\ref{eq:orth}). Recall that the bound in Eq. (\ref{eq:bdnos1}) is about the ME discrimination among quantum states $\{ \rho_{\x}\}_{\x=1}^{N}$ that are given with prior probabilities $\{q_{\x} =p_{\x} / \sum_{\y=1}^{N} p_{\y} \}_{\x=1}^{N}$. All these are summarized as follows.

\begin{lem}
\label{lem:guno}
When quantum states $\{ q_{\x}, \rho_{\x}\}_{\x=1}^{N}$ are prepared as it is shown in Eq. (\ref{eq:ens}) by ensemble steering, the guessing probability is bounded upper by the no-signaling condition,
\bea
P_{\g} \leq \frac{1}{ \sum_{\x=1}^{N} p_{\x}},~\mathrm{with~ equality~ if ~and ~only~ if} ~P(\x | \x,   \{ \sigma_{\x}\}_{\x=1}^{N}  ) = 0, ~\mathrm{for}~\x = 1,\cdots,N, \label{eq:bdnos2}
\eea
where $\{p_{\x}\}_{\x=1}^{N}$ are from Eq. (\ref{eq:ens}). With the measurement postulate, the equality holds when POVM $M_{\x}$ does not respond to $\sigma_{x}$ but only to $\rho_{\x}$ for each $\x$, that is,
\bea
\tr[M_{\x} \sigma_{\x}] = 0\label{eq:bdsig}
\eea
which reproduces the condition in Eq. (\ref{eq:orth}). Then, the upper bound is equal to the guessing probability from quantum theory, e.g. Eq. (\ref{eq:dual}).
\end{lem}

\emph{Proof.} The upper bound can be derived from Eq. (\ref{eq:bdnos1}). We also recall $q_{\x} = p_{\x} / \sum_{\y=1}^{N}$. Since $\sum_{\x=1}^{N} q_{\x} =1$, the upper bound in Eq. (\ref{eq:bdnos2}) is obtained. $\Box$\\

Consequently, two constraints in Eqs. (\ref{eq:ens}) and (\ref{eq:bdsig}) are optimality conditions for probabilities in the ME discrimination. Optimality conditions in Eqs. (\ref{eq:ens}) and (\ref{eq:bdsig}) are equivalent to those in Lemma \ref{lem:opt}: Eq. (\ref{eq:ens}) is equivalent to the symmetry operator in Eq. (\ref{eq:symop}). We also remark that this approach of constraining with the no-signaling principle is valid for generalized probability theories, as long as the steering effect is allowed to do, i.e.  such that different decompositions of an identical ensemble in Eq. (\ref{eq:ens}) can be prepared at a distance.  
 

\subsection{Geometric formulation} 
\label{subsec:geofom}

From different approaches to the ME discrimination, we have shown different forms of the optimality conditions, which are also shown to be equivalent to the KKT conditions. Compared to previously known forms of the conditions in Eqs. (\ref{eq:opt1}) and (\ref{eq:opt2}), the usefulness of expressing the optimality conditions in KKT conditions in Lemma \ref{lem:opt} is that conditions about states (in the state space) and optimal measurement are separated. 

We here interpret the symmetry operator in terms of the quantum state geometry, and put forward to a geometric approach to optimal state discrimination. This shows that optimal discrimination of quantum states, which is supposed to be explained on the level of probability distributions from measurement outcomes, can be explained only with quantum states and their geometry. 

We first recall the optimality conditions in Lemma \ref{lem:opt}: for a problem of ME discrimination, there exists a symmetry operator $K^{\star}$ which characterizes optimal discrimination and directly gives the guessing probability. Once the operator is obtained, the rest is straightforward. First, complementary states $\{r_{\x}^{\star}, \sigma_{\x}^{\star}\}_{\x=1}^{N}$ can be found as $K^{\star} - q_{\x} \rho_{\x}$ from Eq. (\ref{eq:symop}). Since $K^{\star}$ is uniquely determined by given states, this also means the uniqueness of complementary states in the ME discrimination. 

\begin{lem}
\label{lem:comp}
The symmetry operator and complementary states in a problem of ME discrimination are unique. 
\end{lem}

\emph{Proof.} We first show that the symmetry operator $K^{\star}$ is unique for given states $\{q_{\x},\rho_{\x} \}_{\x=1}^{N}$. Suppose that for given states $\{q_{\x},\rho_{\x} \}_{\x=1}^{N}$, there exist two symmetry operators $K^{\star}$ and $\bar{K}^{\star}$ such that both give the same guessing probability i.e. $P_{\g} = \tr[K^{\star}] = \tr[\bar{K}^{\star} ]$ while 
\bea
K^{\star} & = & q_{\x} \rho_{\x}  + r_{\x}^{\star} \sigma_{\x}^{\star},~~ \forall \x, \nonumber \\
\bar{K}^{\star} & = & q_{\x} \rho_{\x}  + r_{\x}^{\star}  \bar{\sigma}_{\x}^{\star},~~ \forall \x, \label{eq:kk}
\eea
with corresponding complementary states $\{ r_{\x}^{\star}, \sigma_{\x}^{\star} \}_{\x=1}^{N}$ and $\{ r_{\x}^{\star}, \bar{\sigma}_{\x}^{\star} \}_{\x=1}^{N}$, respectively. Note that the parameters $\{ r_{\x}^{\star} \}_{\x=1}^{N} $ remain the same in both cases of complementary states: this follows from Eq. (\ref{eq:symop}) that the guessing probability is also given as $P_{\g} = q_{\x} + r_{\x}^{\star}$ for each $\x$. In addition, let us assume that $\{M_{\x}^{\star} \}_{\x=1}^{N}$ and $\{ \bar{M}_{\x}^{\star} \}_{\x=1}^{N}$ are optimal measurements, respectively, such that, $P_{\g} = \tr[K^{\star} ]  = \tr[\sum_{\x} q_{\x} \rho_{\x} M_{\x}^{\star}  ] $ and also $P_{\g} = \tr[\bar{K}^{\star} ]  = \tr[ \sum_{\x} q_{\x} \rho_{\x} \bar{M}_{\x}^{\star}  ]$, or equivalently from the optimality condition in Eq. (\ref{eq:orth}), $r_{\x}^{\star} \tr[\sigma_{\x}^{\star} M_{\x}^{\star} ] = 0$ and $r_{\x}^{\star} \tr[\bar{\sigma}_{\x}^{\star} \bar{ M}_{\x}^{\star} ] = 0$. 

Now, with two equations in Eq. (\ref{eq:kk}), let us compute $\sum_{\x} K^{\star} \bar{M}_{\x}^{\star}$ and $\sum_{\x} \bar{K}^{\star} M_{\x}^{\star}$ as follows, since $\sum_{\x} M_{\x}^{\star} = \sum_{\x} \bar{M}_{\x}^{\star} = I$,
\bea
K^{\star} & = &  \sum_{\x} K^{\star} \bar{M}_{\x}^{\star}   = \sum_{\x} q_{\x} \rho_{\x} \bar{M}_{\x}^{\star}  +  \sum_{\x} r_{\x}^{\star} \sigma_{\x}^{\star} \bar{M}_{\x}^{\star}  =  \bar{K}^{\star}  +  \sum_{\x} r_{\x}^{\star} \sigma_{\x}^{\star} \bar{M}_{\x}^{\star},      \label{eq:pr1} \\
 \bar{K}^{\star} & = & \sum_{\x} \bar{K}^{\star} M_{\x}^{\star}  =  \sum_{\x} q_{\x} \rho_{\x} M_{\x}^{\star}  +  \sum_{\x} r_{\x}^{\star} \bar{\sigma}_{\x}^{\star} M_{\x}^{\star} = K^{\star}  +  \sum_{\x} r_{\x}^{\star} \bar{\sigma}_{\x}^{\star} M_{\x}^{\star} \label{eq:pr2} 
\eea
where we have used the fact that $K^{\star} = \sum_{\x} q_{\x} \rho_{\x} M_{\x}^{\star}$ and $\bar{K}^{\star} = \sum_{\x} q_{\x} \rho_{\x} \bar{M}_{\x}^{\star}$. From Eqs. (\ref{eq:pr1}) and (\ref{eq:pr2}), we have
\bea
&& K^{\star} + \bar{K}^{\star} = \bar{K}^{\star} +  K^{\star} + \sum_{\x} r_{\x}^{\star}  \sigma_{\x}^{\star} \bar{M}_{\x}^{\star} + \sum_{\x} r_{\x}^{\star} \bar{\sigma}_{\x}^{\star} M_{\x}^{\star}   \nonumber \\
&\Rightarrow& \sum_{\x} r_{\x}^{\star}  \sigma_{\x}^{\star} \bar{M}_{\x}^{\star} +  \sum_{\x} r_{\x}^{\star}\bar{\sigma}_{\x}^{\star} M_{\x}^{\star}   =0 \nonumber \\
&\Rightarrow & \sum_{\x} r_{\x}^{\star}  \sigma_{\x}^{\star} \bar{M}_{\x}^{\star} =0 ~\mathrm{and}   ~ \sum_{\x} r_{\x}^{\star} \bar{\sigma}_{\x}^{\star} M_{\x}^{\star}  =0.  \label{eq:pr}
\eea
To conclude Eq. (\ref{eq:pr}), we have recalled that all operators of measurement and complementary states are positive semidefinite. Plugging Eq. (\ref{eq:pr}) to Eqs. (\ref{eq:pr1}) and (\ref{eq:pr2}), we have that $K^{\star} = \bar{K}^{\star} $. This proves that for a set of quantum states the symmetry operator is unique. 

Then, from Eq. (\ref{eq:symop}) and the uniqueness of the symmetry operator, it is shown that complementary states are also uniquely determined, $\forall \x$, $r_{\x}^{\star} \sigma_{\x}^{\star} = K - q_{\x}\rho_{\x} $. $\Box$\\

For obtained complementary states, it is not difficult to find POVM elements satisfying the orthogonality condition in Eq. (\ref{eq:orth}). An optimal POVM for state $\rho_{\x}$ can be found in the kernel of the state $\sigma_{\x}^{\star}$, $\mathcal{K}[\sigma_{\x}^{\star}]  = \mathrm{span} \{ | \psi\rangle : \sigma_{\x}^{\star} |\psi\rangle = 0\}$, i.e. $M_{\x}^{\star} \in \mathcal{K}[\sigma_{\x}^{\star}]$. In doing this, POVM elements should be chosen such that $\mathrm{span}\{ M_{\x}^{\star} \}_{\x=1}^{N} =  \mathrm{span} \{ \rho_{\x}\}_{\x=1}^{N}$ so that the completeness condition $\sum_{\x=1}^{N} M_{\x}=I$ is fulfilled. Note also that it holds, $\mathrm{span} \{  \rho_{\x}\}_{\x=1}^{N} = \mathrm{span} \{ \sigma_{\x}^{\star}\}_{\x=1}^{N}$. If given states $\{\rho_{\x}\}_{\x=1}^{N}$ are linearly independent, optimal POVM elements must be of rank-one \cite{ref:eldarvn}. 

Since complementary states are unique, one can alternatively solve the ME discrimination in terms of complementary states, without referring directly to optimal measurement. This is also because optimal POVMs are generally not unique: for complementary states $\{ \sigma_{\x}^{\star} \}_{\x=1}^{N}$, optimal POVMs $\{ M_{\x}^{\star} \}_{\x=1}^{N}$ specified by the optimality condition $\tr[ \sigma_{\x}^{\star} M_{\x}^{\star} ]$ are not unique \cite{ref:mochon}. From Lemma \ref{lem:opt}, optimal discrimination is solved once those parameters satisfying the KKT conditions. If the underlying geometry of given states is clear, complementary states can be found from given states to discriminate among, in the following way.

Let $\mathcal{P}(\{q_{\x} ,\rho_{\x} \}_{\x=1}^{N})$ denote the polytope of given states constructed in the state space, where each vertex of the polytope is specified by $q_{\x}\rho_{\x}$. The condition of symmetry operator in Eq. (\ref{eq:symop}) can be written as,
\bea
q_{\x} \rho_{\x}  - q_{\y} \rho_{\y} = r_{\y}^{\star} \sigma_{\y}^{\star} - r_{\x}^{\star} \sigma_{\x}^{\star},~\forall~\x,\y=1,\cdots,N. \label{eq:optpol} 
\eea
This means that the unknown polytope of complementary states $\mathcal{P} ( \{ r_{\x}^{\star},\sigma_{\x}^{\star}\}_{\x=1}^{N})$ is congruent to the polytope of given states. Note that, here, we say that two polytopes are congruent if all of vertices of one polytope are identical to those of the other. This already determines the polytope $\mathcal{P} ( \{ r_{\x}^{\star},\sigma_{\x}^{\star}\}_{\x=1}^{N})$ of the states we search for. Then, the optimal discrimination follows by locating $\mathcal{P} ( \{ r_{\x},\sigma_{\x}\}_{\x=1}^{N})$ in the state space such that, together with given polytope $\mathcal{P}(\{q_{\x} ,\rho_{\x} \}_{\x=1}^{N})$, the symmetry operator can be constructed. This is equivalent to the condition that corresponding lines are anti-parallel, see Eq. (\ref{eq:optpol}). An approach to construct the symmetry operator can be done by rewriting the symmetry operator in Eq. (\ref{eq:symop}) as,
\bea
K^{\star} = \frac{1}{N} \sum_{\x=1}^{N} q_{\x} \rho_{\x} + \frac{1}{N} \sum_{\x=1}^{N} r_{\x}^{\star} \sigma_{\x}^{\star}.\label{eq:cm}
\eea
Note an interpretation of $K^{\star}$ that a symmetry operator corresponds to the sum of two centers of two respective polytopes $\mathcal{P}(\{q_{\x} ,\rho_{\x} \}_{\x=1}^{N})$ and $\mathcal{P} ( \{ r_{\x}^{\star},\sigma_{\x}^{\star} \}_{\x=1}^{N})$. Then, given the two congruent polytopes in the state space, a symmetry operator can be obtained by rotating the not-yet-fixed one $\mathcal{P} ( \{ r_{\x},\sigma_{\x}\}_{\x=1}^{N})$ with respect to the fixed one $\mathcal{P} ( \{ q_{\x},\rho_{\x}\}_{\x=1}^{N})$, such that operators from two constructions in Eqs. (\ref{eq:symop}) and (\ref{eq:cm}) are identical. 

To apply the geometric formulation in the above, one should be able to describe the geometry of quantum states in the state space. The difficulty is clearly the lack of a general picture to quantum state space apart from two-dimensional cases, qubit state space.




\section{General structures }
\label{sec:general}

In this section, we show general structures of the ME discrimination: equivalence classes of sets of quantum states, construction of the ME discrimination, and general expressions of the guessing probability. We mainly exploit results shown in the previous section, that for the ME discrimination there always exists a symmetry operator which completely characterizes the optimal discrimination. We first define equivalence classes of sets of quantum states in terms of a symmetry operator. As an approach converse to optimal discrimination for given states, we present a systematic way of constructing a set of quantum states for which the optimal discrimination is immediately known from a given symmetry operator. We then show a general and analytic expression of the guessing probability. 

From now on, unless specified otherwise, for simplicity let $K$ and $\{ r_{\x}, \sigma_{\x} \}$ without $^{\star}$ denote a symmetry operator and complementary states, respectively.


 \subsection{Equivalence classes}

From Lemma \ref{lem:opt}, a symmetry operator gives a complete characterization of optimal parameters in the ME discrimination. Once a symmetry operator is found, the rest to find complementary states and optimal measurements is straightforward. Note that the guessing probability is given by the trace norm of a symmetry operator, see Eq. (\ref{eq:dual}). Therefore, if two different sets of quantum states share an identical symmetry operator, the ME discrimination is characterized in an equivalent way in terms of the identical symmetry operator.

\begin{defn}
\label{def:eqc}
(Equivalence classes)\\ 
Two sets of quantum states, say  $\{q_{\x},\rho_{\x}\}_{\x=1}^{N}$ and $\{{q'}_{\x},{\rho'}_{\x} \}_{\x=1}^{L}$, are equivalent in the ME state discrimination if their symmetry operators are identical up to unitary transformations i.e. identical spectrum. We write two equivalent sets as,  
\bea \{q_{\x},\rho_{\x}\}_{\x=1}^{N} \sim \{{q'}_{\x},{\rho'}_{\x} \}_{\x=1}^{L}, \nonumber 
\eea 
and the equivalence class characterized by a symmetry operator $K$ is denoted by $\mathcal{A}_{K}$. Then, sets of quantum states in the same equivalence class have the same guessing probability. 
\end{defn}

\subsection{Construction of a set of quantum states from a symmetry operator}

In this subsection, conversely to solving a problem of ME discrimination (or equivalently to finding the symmetry operator), we here introduce a systematic way of generating a set of quantum states from a given symmetry operator. This means that, for the generated quantum states, optimal discrimination is already characterized by the symmetry operator. That is, in this way, elements of an equivalent class identified by a symmetry operator are generated. 

The main idea is to exploit the structure of the ME discrimination shown in the subjection \ref{subsec:fun}. Suppose that a symmetry operator is given by $K \in \B (\H)$ which is simply a (bounded) positive operator over a Hilbert space $\H$. As the guessing probability is to be given as $\tr[K]$ at the end (see, Eq. (\ref{eq:dual})), we note that the operator is not larger than the identity operator in the space. To construct a set of quantum states $\{q_{\x}, \rho_{\x} \}_{\x=1}^{N}$ having the operator $K$ as their symmetry operator, the first thing to do is to normalize the operator to interpret it as a quantum state and then make its purification. Let $\K$ denote the operator after normalization, $\K = K/ \tr[K]$. We write its purification as $|\psi_{K}\rangle_{AB} \in  \H \otimes \H$ such that $\K = \tr_{A} | \psi_{K}\rangle_{AB} \langle \psi_{K}|$, and the purification is unique up to local unitary transformations on the $A$ system. 

Then, the next is to construct $N$ two-outcome and complement measurements on the $A$ system, $M^{\x} = \{ M_{0}^{\x}, M_{1}^{\x} \}$ with $M_{0}^{\x} +  M_{1}^{\x} = I$ for $\x=1,\cdots,N$. Since each measurement is complete, the resulting state in the $B$ system on average (i.e. ensemble average) is described by the operator $\K$. Decompositions of the operator $\K$ are determined by the choice of measurement $M^{\x}$ on the $A$ system: let $\rho_{\x}$ ($\sigma_{x}$) denote the state resulted by the detection event appeared in Alice's measurement $M_{0}^{\x}$ ($M_{1}^{\x}$), and we assume that the detection event happens with probability $p_{\x}$ ($1-p_{\x})$. In this way, there are $N$ different decompositions of the operator $\K$,
\bea
\K & = & p_{\x} \rho_{\x} +(1-p_{\x}) \sigma_{\x}, ~\mathrm{for}~\x = 1,\cdots,N \label{eq:ktil}
\eea
where $\{\rho_{x} \}_{x=1}^{N}$ are those states that we are interested to discriminate. Given that a measurement device is prepared to discriminate among these states $\{\rho_{x} \}_{x=1}^{N}$, the {\it a priori} probability that $\rho_{\x}$ is generated is given by $p_{\x} / \sum_{\x=1}^{N} p_{\x}$, which we write by $q_{\x}$. 

The ensemble decomposition in Eq. (\ref{eq:ktil}) corresponds to the case shown in Eq. (\ref{eq:ens}), or also equivalently in Eq. (\ref{eq:symop}), that an identical ensemble is decomposed into $N$ different ways. In this case, the optimality conditions are presented in Eq. (\ref{eq:bdnos2}) which is also equivalent to the optimality condition in Eq. (\ref{eq:orth}). That is, it immediately follows that optimal discrimination of states $\{q_{\x},\rho_{\x} \}_{\x=1}^{\x}$ is characterized: the guessing probability is $P_{\g} = \tr[K]$ from the given operator, and complementary states are $\{r_{\x}, \sigma_{\x}\}_{\x=1}^{N}$ where $r_{\x}$ can be found from the relation in Eq. (\ref{eq:symop}), $r_{\x} = (1-p_{\x}) / \tr[K]$ and $\sigma_{\x}$ results from measurement $M_{1}^{\x}$ in the $A$ system. 

\begin{prop} 
\label{prop:con}
(Construction of equivalence classes)\\
When a symmetry operator $K$ is provided, an element of the equivalent class $\mathcal{A}_{K}$ can be constructed as a set of states $\{ q_{\x},\rho_{\x}\}_{\x=1}^{N}$ if there exist a set of POVM elements $\{ M_{0}^{\x} \}_{\x=1}^{N}$ on the $A$ system of the purification $|\psi_{K}\rangle_{AB}$ such that those states $\{\rho_{\x}\}_{\x=1}^{\x}$ are prepared in the $B$ system with probabilities $\{ p_{\x}\}_{\x=1}^{N}$, respectively, where $p_{\x}= q_{\x} \tr[K] $ for each $\x = 1,\cdots,N$. Then, the other POVM elements $\{ M_{1}^{\x} \}_{\x=1}^{N}$, which fulfill that $M_{0}^{\x} +M_{1}^{\x} = I$ for each $\x$, uniquely find complementary states $\{\sigma_{\x} \}_{\x=1}^{N}$ in the $B$ system.
\end{prop}


Before to proceed, we present an example that shows how the method of constructing a set of quantum states can be applied, when a symmetry operator is given. This is simple but also useful to see how all that have been explained work out. Further examples are also presented in Sec. \ref{sec:qubit}. \\

\emph{Example}. (Equivalence class of a normalized identity operator) Suppose that a symmetry operator is given as the identity operator in a $d$-dimensional Hilbert space, $K = I/d$. Take one of spectral decomposition of the identity, we write it using orthonormal basis $\{ | k \rangle \}_{ k=1}^{d}$: $I = \sum_{k=1}^{N} |k\rangle \langle k|$. Since the symmetry operator is normalized, the guessing probability for quantum states to be constructed is to be $P_{\g}=1$. The purification is the maximally entangled state, $|\psi_{K} \rangle = \sum_{k=1}^{d} |k\rangle |k\rangle /\sqrt{d}$. Applying measurement $\{ M_{0}^{\x} = |\x \rangle \langle \x |, M_{1}^{\x} = I - M_{0}^{\x} \}_{\x=1}^{d}$ on the $A$ system, the following ensemble is prepared in the $B$ system,
\bea
K = \frac{1}{d} |\x \rangle \langle \x| + \frac{d-1}{d} \sigma_{\x},~\mathrm{where}~\sigma_{\x} = \frac{1}{d-1}\sum_{\y \neq \x} |\y\rangle \langle \y |,~\forall~{\x=1,\cdots,d}  \nonumber
\eea
This defines a problem of state discrimination for orthogonal states $\{1/d, |\x\rangle\langle \x| \}_{\x=1}^{N}$. Complementary states are shown in the above, and optimal measurements are also straightforward to find. Thus, it is shown that $\{1/d, |\x\rangle \langle \x| \}_{\x=1}^{N} \in \mathcal{A}_{I/d}$. \\

\subsection{Analytic expression of the guessing probability}
\label{sec:fwork}

In this subsection, we present various expressions of the guessing probability in ME state discrimination. They are equivalent but of different forms depending on how they are derived. The first one based on the optimality conditions corresponds to a quantum analogy of the probabilistic-theoretic measure shown in Eq. (\ref{eq:lem2-1}). The next one is in accordance with other physical theories, ensemble steering on quantum states and the no-signaling principle on measurement outcomes. 

\subsubsection*{Quantum analogy to the probability-theoretic expression}

Let us first present a general form of the guessing probability (for quantum states) in the framework of probability-theoretic measures. We first recall from Lemma \ref{lem:2-1} that the guessing probability about random variable $X$ given $Y$ is generally expressed as the distance of probability $P_{X|Y}$ deviated from the uniform distribution. This holds true in general, no matter what physical systems are applied to mediate between the preparation and the measurement. Then, it remains to determine the form of $d(X|Y)$ once quantum systems are applied.

We recall the KKT conditions of the ME discrimination for $\{q_{\x},\rho_{\x} \}_{\x=1}^{N}$, as follows
\bea
K = q_{\x} \rho_{\x} + r_{\x} \sigma_{\x},~\mathrm{for}~\x = 1,\cdots,N,~\mathrm{with~ complementary~ states}~ \{r_{\x},\sigma_{\x} \}_{\x=1}^{N}.\nonumber
\eea
Using the simple identity that $K = \sum_{\x=1}K/N$, we have
\bea
P_{\g} = \frac{1}{N} \tr[ \sum_{\x} q_{\x} \rho_{\x} + r_{\x} \sigma_{\x}] = \frac{1}{N} + R, ~\mathrm{ where} ~ R= \frac{1}{N} \sum_{\x=1}^{N} r_{\x},\label{eq:gusp}
\eea
This compares to Eq. (\ref{eq:lem2-1}): the distance $d(X|Y)$ corresponds to the dual parameter $R$, which now corresponds to the norm of the ensemble average of complementary states. 

The parameter $R$ in the expression in Eq. (\ref{eq:gusp}) in fact has a geometrical meaning in the state space. Recall from the subsection \ref{subsec:geofom} that each complementary state is determined as a vertex of the polytope that is congruent to the polytope constructed by given states, $\mathcal{P} (\{q_{\x},\rho_{\x} \}_{\x=1}^{N} )$. Then, each state $r_{\x} \sigma_{\x}$ plays the role of the (trace) distance between each vertex $q_{\x} \rho_{\x}$ of the polytope and the symmetry operator, i.e.
\bea
r_{\x} = \tr [ r_{\x} \sigma_{\x} ]= \tr [ K - q_{\x} \rho_{\x} ] = \| K - q_{\x} \rho_{\x}  \|_{1}, \nonumber
\eea
where note that $K \geq q_{\x} \rho_{\x}$ for all $\x=1,\cdots,N$, from the dual problem in Eq. (\ref{eq:dual}). 

\begin{prop} 
\label{lem:qguess}
The guessing probability for quantum states $\{ q_{\x} ,\rho_{\x} \}_{\x=1}^{N}$ constructed from a symmetry operator $K$ is,
\bea
P_{\g} = \frac{1}{N} + R (K || \{ q_{\x} ,\rho_{\x} \}_{\x=1}^{N} ),~ \mathrm{with}~R (K || \{ q_{\x} ,\rho_{\x} \}_{\x=1}^{N} ) = \frac{1}{N} \sum_{x=1}^{N} \| K  - q_{\x} \rho_{\x}  \|_{1},  \label{eq:qguess}
\eea
where $R (K || \{ q_{\x} ,\rho_{\x} \}_{\x=1}^{N} )$ shows the averaged trace distance of given states $\{ q_{\x} ,\rho_{\x} \}_{\x=1}^{N}$ deviated from their symmetry operator $K$, see also Eq. (\ref{eq:lem2-1}) in Lemma \ref{lem:2-1} for comparison.
\end{prop}

\subsubsection*{General expression for probabilistic and physical models}

The next form of the guessing probability (for discrimination among states $\{ q_{\x},\rho_{\x} \}_{\x=1}^{N}$) is obtained by the compatibility between ensemble steering on quantum states and the no-signaling condition on probability distributions of measurement outcomes. Then, as it is shown in Lemma \ref{lem:guno}, the guessing probability is given by 
\bea
P_{\g} = \frac{1}{p_1 + \cdots + p_{N}}, ~\mathrm{with}~ p_{\x} = q_{\x} P_{\g},~\mathrm{for}~\x=1,\cdots,N \label{eq:guessphy}
\eea
since the bound in Eq. (\ref{eq:bdnos2}) is tight. From the ensemble steering shown in Eq. (\ref{eq:ens}), the guessing probability can be interpreted as follows. Each parameter $p_{\x}$ of $\{ p_{\x}\}_{\x=1}^{N}$ in the above shows the probability that a party can prepare a quantum state $\rho_{\x}$ to the other party at a distance via the ensemble steering. Therefore, the guessing probability corresponds to the maximal average probability to prepare a set of chosen quantum states to a distant party via the ensemble steering.

\begin{thm} 
(Steerability and state discrimination)\\
While ensemble steering is allowed within quantum theory, the steerability of a quantum state in the ensemble is generally proportional to the guessing probability for those states prepared in the ensemble. 
\end{thm}

\subsubsection*{Simplification for equal prior probabilities}

Finally, for the ME discrimination for $\{ q_{\x},\rho_{\x} \}_{\x=1}^{N}$, a huge simplification can be made when prior probabilities are equal i.e. $q_{\x}=1/N$ for all $\x$. The simplification is shown in the expression of $R$ in Eq. (\ref{eq:qguess}), that it holds $r_{\x} = r_{\y}$ for all $\x,\y=1,\cdots,N$ if $q_{\x} =1/N$. Note also that since the symmetry operator in Eq. (\ref{eq:symop}) shows that $P_{\g} = \tr[K] = q_{\x} + r_{\x} = q_{\y} + r_{\y}$ $\forall \x,\y$, and now since $q_{\x} = q_{\y}=1/N$, it holds that $r_{\x} = r_{\y}$ for all $\x,\y$. From Lemma \ref{lem:qguess}, this means that distances of given states from the symmetry operator are all equal. Thus, for convenience, let us write 
\bea 
r:= r_{\x},~\forall~\x=1,\cdots,N,~~\mathrm{and~then,~} r = R(K \|  \{ 1/N ,\rho_{\x} \}_{\x=1}^{N} ) . \label{eq:r1}
\eea
The problem of ME discrimination becomes even simpler. There is only a single parameter $r$ to find, to solve optimal quantum state discrimination for the uniform prior.

Applying the geometric formulation shown in the subsection \ref{subsec:geofom}, the geometrical meaning of the parameter $r$ in Eq. (\ref{eq:r1}) can be found in a simple way. From the KKT condition in Eq. (\ref{eq:symop}), we have the following relation, see also Eq. (\ref{eq:optpol}),
\bea
\frac{1}{N} \rho_{\x} - \frac{1}{N} \rho_{\y} = r \sigma_{\y} - r \sigma_{\x},~\mathrm{and~then},~ r = \frac{ \|  \frac{1}{N} \rho_{\x}  - \frac{1}{N} \rho_{\y} \|_{1} }{ \| \sigma_{\x} - \sigma_{\y}  \|_{1} } \label{eq:r}
\eea
Then, the parameter $r$ corresponds to the ratio between two polytopes in the state space, the given one $\mathcal{P}(\{ \frac{1}{N},\rho_{\x} \}_{\x=1}^{N} )$ constructed from given states and the other one $\mathcal{P}(\{ \sigma_{\x} \}_{\x=1}^{N} )$ only of complementary states.

\begin{prop}
\label{prop:unidis}
The guessing probability for quantum states given with the uniform prior is determined by only a single parameter as
\bea
P_{\g} = \frac{1}{N} + r, \label{eq:guessr}
\eea
where $r$ is the ratio between two polytopes, one from given states and the other from complementary states. 
\end{prop}


\section{Examples: solutions in qubit state discrimination}
\label{sec:qubit}

In this section, we apply general results shown so far to an arbitrary set of qubit states. Since qubits are the unit of quantum information processing, results to be presented here are not only of theoretical interest to characterize quantum capabilities of state-discriminateion-based tasks, but also useful for practical applications. Note that, for ME discrimination for qubit states, general solutions are known for two qubit states. For more than two states, otherwise, optimal discrimination is known for qubit states containing some symmetry such as geometrically uniform structure. In the recent, there has been analytical approaches to qubit state discrimination, one exploiting the dual problem in Eq. (\ref{eq:dual}) \cite{ref:terhal} and the other from the complementarity problem that analyzing the KKT conditions in Eqs. (\ref{eq:symop}) and (\ref{eq:orth}) \cite{ref:bae}. In the latter, an analytic method is provided to solve ME discrimination for any set of qubit states with the uniform prior. In the following subsections, we present optimal discrimination of qubit states when qubit states may not contain any symmetry among them. That is, we present symmetry operator, complementary states, and optimal measurement.

Let us collect general results to be used for qubit state discrimination. As for qubit states, a useful and clear geometric picture is present with the Bloch sphere, in which the state geometry is also clear with the Hilbert-Schmidt distance. The other useful fact is that for qubit states, the Hilbert-Schmidt distance is proportional to the trace distance. From this, the geometry formulation in the subsection \ref{subsec:geofom} can also be applied.
\begin{lem}
For qubit states, the Hilbert-Schmidt and the trace distances, denoted by $d_{HS}$ and $d_{T}$ respectively, are related by a constant as follows, 
\bea
d_{HS}(\rho,\sigma) = \sqrt{2} ~d_{T} (\rho,\sigma), \label{eq:dis}
\eea
for any qubit states $\rho$ and $\sigma$.
\end{lem}
The lemma is useful when the trace distance between qubit states is computed via the geometry in the Bloch sphere: once the actual distance measure in the Bloch sphere is computed in the Hilbert-Schmidt norm via the geometry, from Lemma in the above the distance can be converted to the trace norm.

The next useful fact is the orthogonality relation. In the two-dimensional Hilbert space, if two non-negative and Hermitian operators $O_1$ and $O_2$ fulfill the orthogonality condition $\tr[O_1 O_2]=0$, see Eq. (\ref{eq:orth}), the only possibility is that two operators are of rank-one and, once normalized, they are a resolution of the identity operator. That is, if $O_{1} \propto |\phi \rangle \langle \phi |$ to satisfy the orthogonality, then the other is uniquely determined $O_{2} \propto |\phi^{\perp} \rangle \langle \phi^{\perp}|$, since in the two-dimensional space, the resolution of the identity is given as $I = |\phi\rangle\langle\phi| + |\phi^{\perp}\rangle\langle\phi^{\perp}|$ for any vector $|\phi\rangle$. This means that as long as the optimal strategy in state discrimination is not the application of the null measurement for some state (i.e. $M_{\x}=0$ for some $\x$), optimal POVM elements and complementary states, that would fulfill the optimality condition in Eq. (\ref{eq:orth}) each other, must be of rank-one and uniquely determined from the aforementioned relation in the above. All these are summarized in the following, exploiting the optimality condition in Eq. (\ref{eq:orth}).

\begin{lem}
\label{lem:optqdis}
The followings are conditions for optimal measurement and complementary states in optimal discrimination of qubit states. 
\begin{enumerate}
\item If complementary states are not pure states, then the optimal strategy to have a minimal error is to apply the null-measurement, i.e. no-measurement for some states gives an optimal strategy  \cite{ref:hunter} \cite{ref:bae}.
\item For cases where optimal strategy is not the null measurement, optimal POVM elements are of rank-one \cite{ref:terhal} \cite{ref:bae} and uniquely determined by complementary states that are also of rank-one \cite{ref:bae}.
\end{enumerate}
\end{lem}

\emph{Proof}. $1$.  Let us decompose a complementary state $\sigma_{\x}$ in the following way $\sigma_{\x}= s_{\x} |\varphi_{\x}\rangle \langle \varphi_{\x}| + (1-s_{\x}) |\varphi_{\x}^{\perp}\rangle \langle \varphi_{\x}^{\perp}|$ with some $|\varphi_{x}\rangle$ and $|\varphi_{x}^{\perp}\rangle$ in the two-dimensional space. One can always find two orthogonal vectors to decompose a qubit state in this way. Suppose that $s_{\x} > 0$ so that $\sigma_{\x}$ is not of rank-one. Then, suppose that there exists a POVM element $M_{\x}$ such that $\tr[M_{\x} \sigma_{\x}]=0$. This means that $\langle \varphi_{\x}| M_{\x} | \varphi_{\x} \rangle = \langle \varphi_{\x}^{\perp}| M_{\x} | \varphi_{\x}^{\perp} \rangle =0$. Since $I = |\varphi_{\x} \rangle \langle \varphi_{\x}| + |\varphi_{\x}^{ \perp} \rangle \langle \varphi_{\x}^{\perp}|$, it follows that $\tr[M_{\x}] = \tr[M_{\x} I] =0$. Since $M_{\x}$ is non-negative, we have $M_{\x}=0$, i.e. no-measurement. \\
$2$. Now it is clear that, when two positive operator e.g. $\sigma_{\x}$ and $M_{\x}$ fulfill the orthogonality relation, both of them must be of rank-one. Once a complementary state is given, the only choice for its corresponding POVM element which satisfies the orthogonality is that the POVM is in the kernel of the complementary state $\mathcal{K}[\sigma_{\x}]$. Since $\dim\H=2$ and $\sigma_{\x}$ is of rank-one, it follows that $\dim\mathcal{K}[\sigma_{\x}]=1$. Thus, the POVM element is also of rank-one and uniquely determined in the one-dimensional kernel space of a complementary state. $\Box$\\

Lemma shown in the above applies directly to the geometric formulation presented in the subsection \ref{subsec:geofom} for qubit states. Recall that most of complementary states for which the optimal measurement is not the null POVM element are of rank-one, that is, pure states. This means that complementary states lie at the surface of the Bloch sphere, and the polytope of them $\mathcal{P}(\{ \sigma_{\x}\}_{\x=1}^{N} )$ is therefore the maximal within the Bloch sphere. The shape of the polytope $\mathcal{P}(\{ r_{\x},\sigma_{\x}\}_{\x=1}^{N} )$ is also known to be congruent to the polytope $\mathcal{P}(\{ q_{\x},\rho_{\x}\}_{\x=1}^{N} )$ of given states, see the formulation in the subsection \ref{subsec:geofom}.

Furthermore, let us recall the followings. For the uniform prior probabilities $q_{\x}=1/N$, as it is shown in Eq. (\ref{eq:guessr}) there is only a single parameter to find, in order to solve the optimal discrimination. The parameter is expressed by $r$ in Eq. (\ref{eq:guessr}), corresponding to the ratio between two polytopes $\mathcal{P}(\{ \sigma_{\x}\}_{\x=1}^{N} )$ and  $\mathcal{P}(\{ q_{\x}=1/N,\rho_{\x}\}_{\x=1}^{N} )$, see the expression in Eq. (\ref{eq:r}). Note also that $r = r_{\x}$ for all $\x=1,\cdots,N$ when finding the complementary states $\{r_{\x},\sigma_{\x} \}_{\x=1}^{N}$, as it is shown in Eq. (\ref{eq:r1}). 

\begin{lem}
\label{lem:4-3}
The guessing probability for qubit states given with the uniform prior probabilities is in the following form, 
\bea
P_{\g} = \frac{1}{N} + r\nonumber
\eea
with the parameter $r$, the ratio between two polytopes in the state space, $\mathcal{P}(\{ q_{\x}=1/N, \rho_{\x}\}_{\x=1}^{N} )$ constructed from given states and its similar polytope that is also maximal in the Bloch sphere.
\end{lem}

\emph{Proof}. Two polytopes $\mathcal{P}(\{ q_{\x}=1/N, \rho_{\x}\}_{\x=1}^{N} )$ and $\mathcal{P}(\{ r ,\sigma_{\x}\}_{\x=1}^{N} )$ are congruent, and therefore two polytopes $\mathcal{P}(\{ q_{\x}=1/N,\rho_{\x}\}_{\x=1}^{N} )$ and $\mathcal{P}(\{ \sigma_{\x}\}_{\x=1}^{N} )$ are similar with the ratio $r$. For qubit state discrimination, from Lemma \ref{lem:optqdis} it holds that most complementary states are of rank-one lying at the surface of the Bloch sphere, and hence $\mathcal{P}(\{ \sigma_{\x}\}_{\x=1}^{N} )$ is the maximal in the Bloch sphere. Therefore, the ratio $r$ is given from $\mathcal{P}(\{ q_{\x}=1/N,\rho_{\x}\}_{\x=1}^{N} )$ and its similar and maximal one $\mathcal{P}(\{ \sigma_{\x}\}_{\x=1}^{N} )$. This completes the proof. $\Box$

In what follows, we apply all these results to qubit state discrimination. We also write qubit states using Bloch vectors as, $\rho_{\x} = \rho(\v_{\x}) = \frac{1}{2}(I + \v_{\x} \cdot \vec{\sigma})$ where $\vec{\sigma} = (X, Y, Z)$ are Pauli matrices. 

\subsection{Two states}

When two quantum states are given, the state space spanned by them is effectively two-dimensional. Consequently, discrimination of them is equivalently restricted to two qubit states. This does not lose any generality. Then, for two-state discrimination $\{ q_{\x},\rho_{\x}\}_{\x=1}^{2}$, optimal discrimination was shown in Ref. \cite{ref:hel},
\bea
P_{\g} =  \frac{1}{2}(1+ \| q_1 \rho_{1} - q_{2}\rho_{2} \|_1 ). \label{eq:hb}
\eea
We now reproduce the result using the geometric formalism presented in the subsection \ref{subsec:geofom}.


\begin{figure}[t]
\begin{center}
\includegraphics[width=5in]{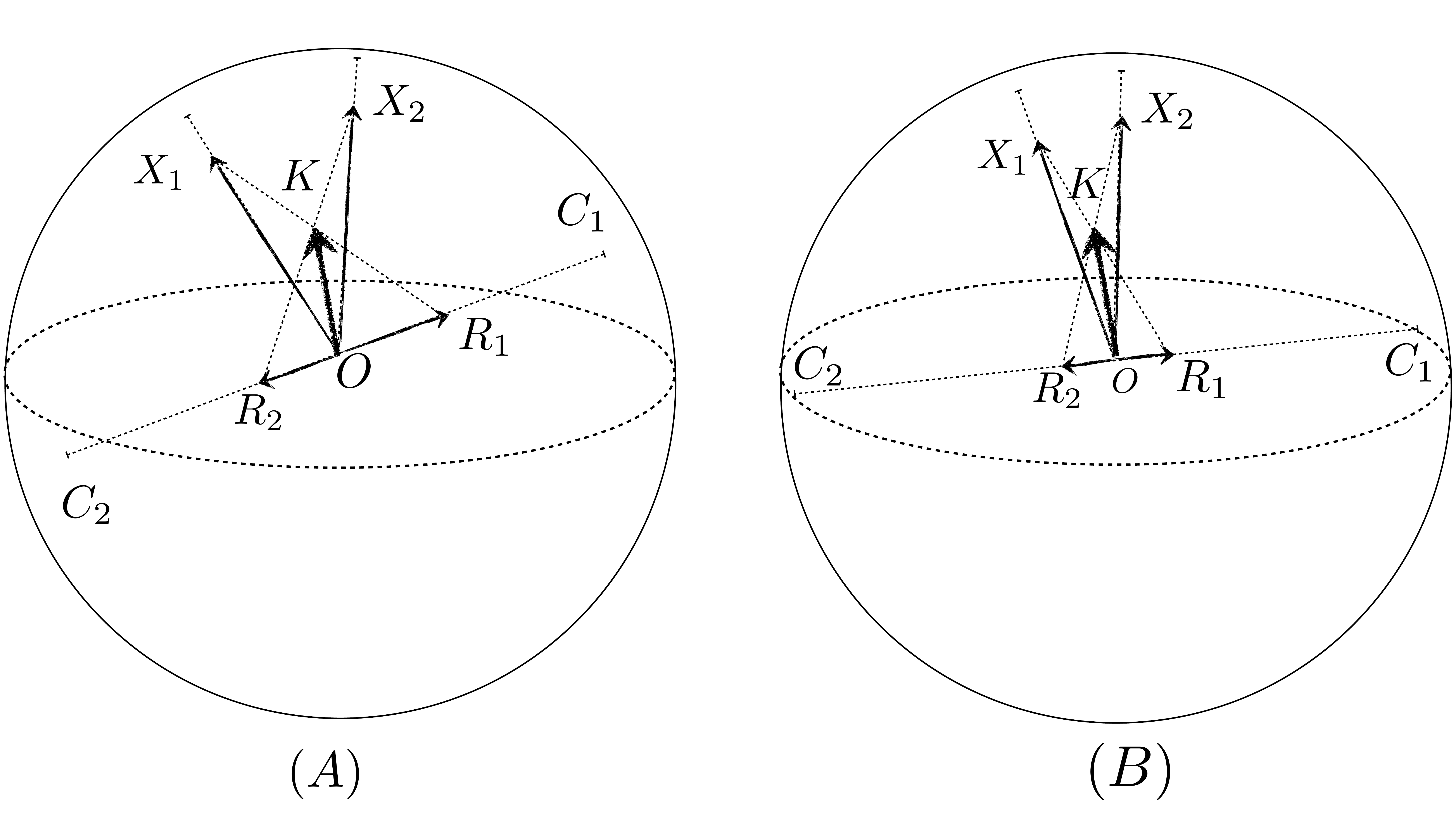}
\caption{(Discrimination of two states $\{ q_{\x},\rho_{\x} \}_{\x=1}^{2}$) In both figures $(A)$ and $(B)$, given states $q_{1}\rho_{1}$ and $q_{2}\rho_{2}$ are depicted by two lines $OX_{1}$ and $OX_{2}$, respectively. The line $OK$ corresponds to the symmetry operator, which is the same in both figures and therefore the guessing probability for two cases is the same, i.e. two cases $(A)$ and $(B)$ are in the same equivalence class. From the geometric formation in the subsection \ref{subsec:geofom}, complementary states can be found as a polytope congruent to the given polytope $X_1 X_2$ of given states. From Lemma \ref{lem:optqdis}, complementary states are on the surface. Therefore, $R_1 R_2$ is the polytope congruent to $X_1X_2$, and then $OC_{1}$ and $OC_{2}$ are complementary states $\sigma_{1}$ and $\sigma_{2}$ respectively, from which optimal POVM elements for states $\rho_1$ and $\rho_2$ are $OC_{1}$ and $OC_{2}$ respectively. The KKT condition in Eq. (\ref{eq:symop}) holds as $OK = OX_1 + OR_1 = OX_2 + OR_2$, and the other in Eq. (\ref{eq:orth}) $OC_1 \perp OC_2$.  }
\label{fig:two state}
\end{center}
\end{figure}


The optimal discrimination is obtained if complementary states are found, since complementary states give the symmetry operator that gives the guessing probability, and also defines optimal measurements, see Lemma \ref{lem:opt}. Thus, we now show how to find complementary states from given states. We first rewrite Eq. (\ref{eq:optpol}) for the two states,
\bea
q_{1}\rho_1 - q_{2} \rho_2 = r_2\sigma_2 - r_1 \sigma_{1}, \label{eq:twocon}
\eea
where $\{ r_{\x},\sigma_{\x} \}_{\x=1}^{2}$ are complementary states to find. Recall Lemma \ref{lem:optqdis}, complementary states are of rank-one and thus lie at the surface of the Bloch sphere. Then, Eq. (\ref{eq:twocon}) in the above means i) two polytopes (lines in this case) by given states and by complementary states, respectively, are congruent and also parallel. Thus, given the line defined by $q_{1}\rho_1 - q_{2} \rho_2$ in the Bloch sphere, two complementary states can be found by finding a diameter (since they are pure states) such that the diameter is parallel to the given line, see Fig.\ref{fig:two state}. Two parameters $r_1$ and $r_2$ can be found to satisfy the relation in Eq. (\ref{eq:twocon}). Then, the symmetry operator is found as $q_{\x}\rho_{\x} + r_{\x} \sigma_{\x}$ for $\x=1,2$, and optimal measurements are also defined in the diameter of complementary states as $M_{1}\propto \sigma_2$ and $M_{2} \propto \sigma_1$.

From the geometric method, the guessing probability can be explicitly written as follows. Note that, from Eq. (\ref{eq:twocon}), i) $\tr[ q_1 \rho_1 - q_2 \rho_2] = q_1 - q_2 $ and ii) $r_1 + r_2 = \| r_1 \sigma_{1} - r_2 \sigma_2 \|$ since $\tr[\sigma_1 \sigma_2] =0$ i.e. two complementary states are found to be orthogonal, and consequently we have that
\bea
i) ~ r_1 - r_2 &=& q_2 - q_1 \nonumber\\
ii)~ r_1 + r_2 & = & \|  r_1 \sigma_{1} - r_2\sigma_2\|_1 = \| q_{1}\rho_1 - q_{2} \rho_2 \|_1. \nonumber 
\eea
Thus, for $\x=1,2$
\bea
r_{\x} = \frac{1}{2} (\| q_1 \rho_1 - q_2 \rho_2\| + (-1)^{\x}(q_{1} - q_{2})  )
\eea
Thus, from the general expression shown in Eq. (\ref{eq:gusp}) for $N=2$, the guessing probability is 
\bea 
P_{\g} & = & \tr[K] = \frac{1}{2} + \frac{1}{2} ( r_1 +r_2) =  \frac{1}{2} (1 + \| q_1 \rho_1 - q_2 \rho_2\|),~\mathrm{with}, \nonumber \\
K & = & \frac{1}{2} (q_1 \rho_1 + q_2 \rho_2) + \frac{1}{2} (r_1 \sigma_1 + r_2 \sigma_2) \label{eq:k2}
\eea
Thus, it is shown that the Helstrom bound is reproduced.

In particular, when $q_{1} =q_{2}=1/2$, from Proposition \ref{prop:unidis} it follows that $r = r_1 = r_2 = \|\rho_{1} - \rho_{2} \|/2$. Then, from the expresso in Eq. (\ref{eq:k2}), the symmetry operator has an even simpler form, 
\bea
K = \frac{1}{2}(\frac{1}{2}\rho_{1} + \frac{1}{2}\rho_2) + \frac{1}{2} (rI). \nonumber
\eea 
since two complementary states correspond to the diameter and their average is simply proportional to the identity. 

In this case, let us also show how the formulation in Lemma \ref{lem:4-3} can be applied. First, note again that the polytope constructed by given states is the line $q_1 \rho_1 - q_2 \rho_2$ in the Bloch sphere. Since it is a line, the largest polytope similar to the the line is clearly a diameter that has the length $2$ in trace norm. The parameter $r$ in Eq. (\ref{eq:guessr}) that we look for is the ratio between two lines,
\bea
\| \frac{1}{2}\rho_1 -\frac{1}{2}\rho_2 \| : 2 = r :1,~\mathrm{and~thus}~ r= \frac{1}{2} \| \frac{1}{2}\rho_1 -\frac{1}{2}\rho_2 \|  \nonumber
\eea
With the above, the guessing probability is $P_{\g} = 1/2+r$, which reproduces the Helstrom bound in Eq. (\ref{eq:hb}) when $q_1 = q_2=1/2$.

\subsection{Three states}

We now move to cases of three qubit states. For three states, no general solution for the ME discrimination has been known to date apart from specific cases when they are symmetric. Here, for three states given with equal prior probabilities, we apply Lemma \ref{lem:4-3} and show how to find optimal discrimination.

\subsubsection{Three-state example I: isosceles triangles}

\begin{figure}[t]
\begin{center}
\includegraphics[width=5in]{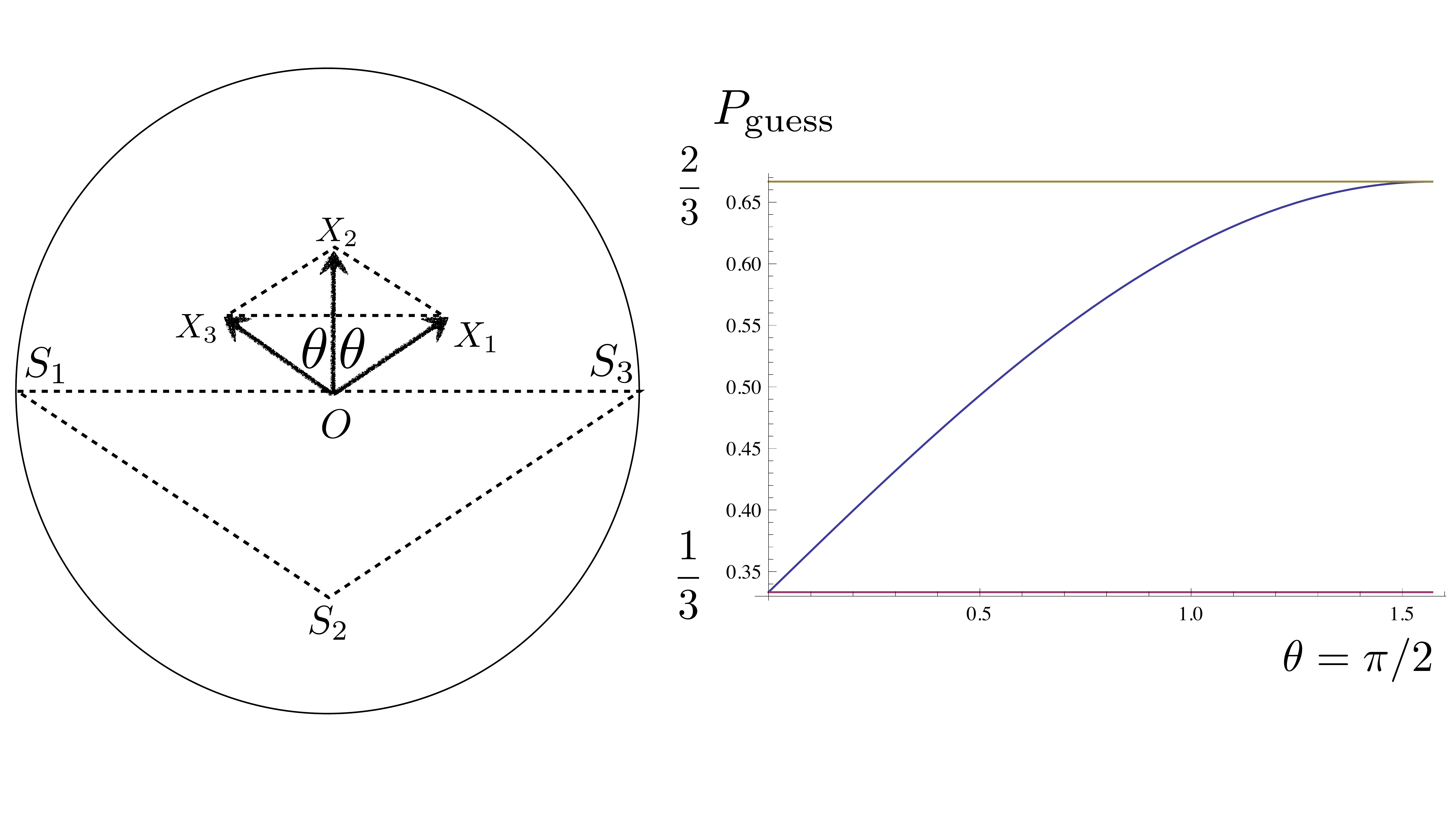}
\caption{Discrimination of three qubit states $\{q_{\x}=1/3,\rho_{\x} = |\psi_{\x}\rangle \langle \psi_{\x}| \}_{\x=1}^{3}$ in a half-plane is shown. It is straightforward to generalize this to three qubit states having the same purity. In the half-plane, each state $\rho_{\x}/3$ corresponds to line $OX_{\x}$ for $\x=1,2,3$. The three states are defined such that two states $\rho_{1}$ and $\rho_{3}$ are equally distant from state $\rho_{2}$, and the distance is specified by angle $\theta$. Then, the polytope of given states corresponds to the triangle $X_1 X_2 X_3$, an isosceles triangle. Since they are given with equal probabilities, Lemma \ref{lem:4-3} is applied as follows. The largest triangle similar to $X_1 X_2 X_3$ is given by  $S_1 S_2 S_3$. The ratio between the two triangles is, $r = (\sin\theta ) / 3$. Thus, the guessing probability is $P_{\g} = (1+\sin\theta)/3$. Note that $OS_{\x}$ for $\x=1,2,3$ correspond to complementary states. Since the complementary state $OS_{2}$ is not pure, optimal POVM for the state $\rho_{2}$ is the null measurement, $M_{2}=0$.} 
\label{fig:three1}
\end{center}
\end{figure}

We first suppose that three pure states lying at a half-plane, and they are given with equal prior probabilities $1/3$, i.e. $\{ 1/3,\rho_{\x}= | \psi_{\x} \rangle \langle \psi_{\x}| \}_{\x=1}^{3}$. In particular, we suppose that the polytope constructed by the three states forms an isosceles triangle on a half plane of the Bloch sphere. For convenience, we parameterize three states as follows, for some $\theta_{0}$,
\bea
|\psi_1\rangle & = & \cos \frac{1}{2}(\theta_0 + \theta ) |0\rangle + \sin \frac{1}{2} (\theta_0 + \theta) |1\rangle \nonumber\\
|\psi_2\rangle & = & \cos \frac{1}{2} \theta_0 |0\rangle + \sin \frac{1}{2}\theta_0 |1\rangle \nonumber\\
|\psi_3\rangle & = & \cos \frac{1}{2} (\theta_0 - \theta ) |0\rangle + \sin \frac{1}{2} (\theta_0 - \theta) |1\rangle, \nonumber\\
\eea
where let us suppose $\theta\in[0,\pi ]$, see Fig. \ref{fig:three1}.


We now apply Lemma \ref{lem:optqdis} to discrimination among these three states, referring to the geometry shown in Fig. \ref{fig:three1}. The maximal polytope which is similar to the given one $X_1 X_2 X_3$ is $S_1 S_2 S_3$. There are many ways of putting the maximal polytope in the plane, however, it should be put as it is shown in the figure to fulfill the optimality condition Eq. (\ref{eq:optpol}), or Eq. (\ref{eq:symop}). Let us first compute the guessing probability. We have to find the ratio $r$ between two polytopes. It corresponds to the ratio, $X_1 X_3/ S_1 S_3$, which is $r = (\sin \theta)/3$. Therefore, the guessing probability is
\bea
P_{\g} = \frac{1}{3} + \frac{1}{3} \sin \theta. \label{eq:three1g}
\eea
In Fig. \ref{fig:three1}, the guessing probability for the three states is plotted as the angle $\theta$ varies. One can easily find that for $\theta\geq \pi/2$, the ratio is $r=1/3$ and thus the guessing probability is, $2/3$.

Let us then find optimal measurement. Note that $OS_1$, $OS_2$, and $OS_3$ correspond to complementary states, $\sigma_1$, $\sigma_{2}$, and $\sigma_{3}$, respectively. It is shown that $\sigma_{2}$ is not a pure state, i.e. not of rank-one. This means that, to fulfill the optimality condition in Eq. (\ref{eq:orth}), the optimal POVM element is the null measurement, i.e. $M_{2}=0$, no-measurement on the state. The other complementary states can be written explicitly as follows,

\bea
\sigma_{1} = | \varphi_1 \rangle \langle \varphi_1|,&& |\varphi_1\rangle = \cos (\frac{\theta_0}{2} - \frac{\pi}{4}) |0\rangle + \sin (\frac{\theta_0}{2} - \frac{\pi}{4}) |1\rangle \\
\sigma_{3} = | \varphi_3 \rangle \langle \varphi_3|,&& |\varphi_3\rangle = \cos (\frac{\theta_0}{2} + \frac{\pi}{4}) |0\rangle + \sin (\frac{\theta_0}{2} + \frac{\pi}{4})|1\rangle \nonumber
\eea
and optimal POVM elements are,
\bea
M_{1} = | \varphi_{3} \rangle \langle \varphi_3|, ~M_{2}=0,~M_{3} = | \varphi_{1} \rangle \langle \varphi_1|,~~\mathrm{so~that}~ \sum_{\x=1}^{3} M_{\x} = I. \nonumber
\eea
One can see that with these POVMs, the guessing probability is obtained 
\bea 
P_{\g}= \frac{1}{3} \tr[ \rho_{1} M_{1}] + \frac{1}{3} \tr[ \rho_{2} M_{2}] + \frac{1}{3} \tr[ \rho_{3} M_{3}]  = \frac{1}{3} + \frac{1}{3} \sin\theta \nonumber
\eea
as it is shown in Eq. (\ref{eq:three1g}). Optimal POVM elements show that for discrimination among the three state given with probability $1/3$, the optimal strategy corresponds to the measurement setting where the device responds only to two most distant states out of the three.

\subsubsection{Three-state example II: geometrically uniform states}

The example shown in the previous can be extended to the so-called geometrically uniform states \cite{ref:eldar} \cite{ref:mirror2}, which correspond to the case that $\theta = 2\pi/3$. It is straightforward to exploit the method shown in the above from Lemma \ref{lem:optqdis}, and then, the guessing probability is $2/3$. In fact, as it is to be shown, the guessing probability is found, $P_{\g} =2/3$ for any three pure (qubit) states given with equal probabilities if the polytope of them in the Bloch sphere contains the origin.

\subsubsection{Three-state example III: arbitrary triangles on a half-plane}

\begin{figure}[t]
\begin{center}
\includegraphics[width=5in]{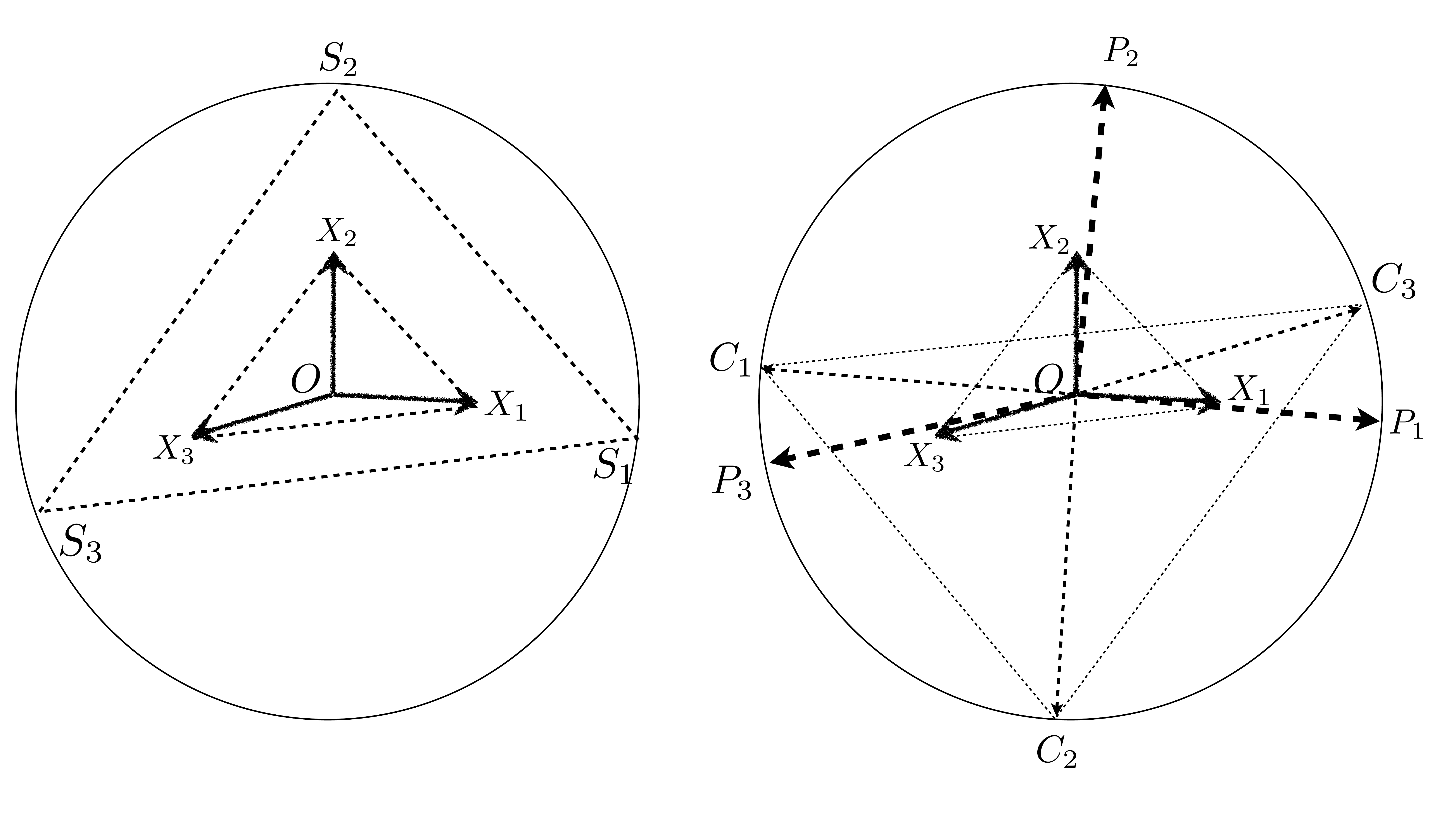}
\caption{ A general method of applying Lemma \ref{lem:4-3} to arbitrarily given three qubit states is shown. Note that prior probabilities are equal, $1/3$, and given states correspond to $OX_{\x}$ for $\x=1,2,3$, and the polytope of given states is $X_1 X_2 X_3$. To make the explanation simple, we suppose that a maximal triangle similar to $X_1 X_2 X_3$, which is $S_1 S_2 S_3$, is covered by the Bloch sphere. In case one of them does not lie at the surface, one can proceed as it is done in Fig. \ref{fig:three1}. Then, the ratio between the two triangles can give the guessing probability. Then, complementary states can be found by rotating $S_1 S_2 S_3$ such that the optimality condition in Eq. (\ref{eq:r}) is fulfilled. Then, the polytope of complementary states $C_1 C_2 C_3$ is defined, where one can see that $X_{\x}X_{\y}$ is parallel with $C_{\y}C_{\x}$ for $\x$ and $\y$. Optimal measurement can also be found as $OP_{\x}$, $\x=1,2,3$, since $OP_{\x}$ is opposite to $OC_{\x}$ from the optimality condition in Eq. (\ref{eq:orth})}
\label{fig:three3}
\end{center}
\end{figure}

We now consider a set of three pure states given with equal probabilities $1/3$, and suppose that the polytope constructed by the three states forms an arbitrary triangle in a half-plane of the Bloch sphere: $\{1/3, \rho_{\x} = |\psi_{\x} \rangle \langle \psi_{\x} | \}_{\x=1}^{3}$. Contrast to the example isosceles triangles shown in the above, we also suppose that the polytope of given states contains the origin of the Bloch sphere, see Fig. \ref{fig:three3}.

To apply the geometric formulation in Lemma \ref{lem:4-3}, we also refer to Fig. \ref{fig:three3}. Note that three states $\rho_{\x}/3$ for $\x=1,2,3$ correspond to $OX_{\x}$ $\x=1,2,3$, respectively, and thus the ploytope of given states is the triangle $X_1 X_2 X_3$. Then, the next to do is to find a maximal polytope similar to $X_1 X_2 X_3$ within the Bloch sphere. A triangle $S_1 S_2 S_3$ in Fig. \ref{fig:three3} can be a possibility. The ratio $r$ corresponds to, for instance, $X_1 X_2 / S_1 S_2$, which is also equal to $OX_1 / OS_1$: therefore $r=1/3$, and then the guessing probability is 
\bea
P_{\g} = \frac{1}{3} + r = \frac{2}{3}. \nonumber
\eea
Note that this is the guessing probability for many cases of three pure states. This can be generalized to cases where three states have the same purity, as follows.

\begin{rem}
Suppose that three qubit states having an equal purity denoted by $f$ are given with equal prior probabilities $1/3$. Here, the purity corresponds to the norm of a Bloch vector. Then, if the polytope of the three states in the Bloch sphere contains the origin, the guessing probability is
\bea
P_{\g} = \frac{1}{3} + \frac{1}{3}f. \nonumber
\eea
This is independent to other details of given quantum states, e.g. angles between them. 
\end{rem}

The triangle $S_1 S_2 S_3$ in Fig. \ref{fig:three3} does not show complementary states yet, since the optimality condition in Eq. (\ref{eq:optpol}), or equivalently Eq. (\ref{eq:symop}), is not fulfilled. Therefore, one has to put or rotate $S_1 S_2 S_3$ to $C_1 C_2 C_3$ in Fig. \ref{fig:three3}, so that Eq. (\ref{eq:optpol}) holds true: $X_{\x}X_{\y} \parallel C_{\y} C_{\x}$ for all $\x,\y=1,2,3$, where $\parallel$ means that two lines are parallel. Then, all optimality conditions are satisfied. 

Complementary states are then those states corresponding to $OC_{\x}$ for $\x=1,2,3$. It also follow that optimal POVM elements are $OP_{\x}$ for $\x=1,2,3$ as each of them is orthogonal to its corresponding complementary state, see the optimality condition in Eq. (\ref{eq:orth}). Thus, optimal POVM elements are $\{M_{\x} \propto OP_{\x}\}_{\x=1}^{3}$. Since the convex hull $P_1 P_2 P_3$ contains the origin, it is also straightforward to have the completeness, $\sum_{\x=1}^{3} M_{\x} =I$. Here, it is noteworthy that optimal POVMs can be in vectors which are not parallel to corresponding states: in Fig. \ref{fig:three3}, we found that $OX_{\x} \nparallel OP_{\x}$ for some $\x$.

\subsubsection{Three-state example IV: arbitrary triangles }
\label{subsec:three4}

\begin{figure}[t]
\begin{center}
\includegraphics[width=5in]{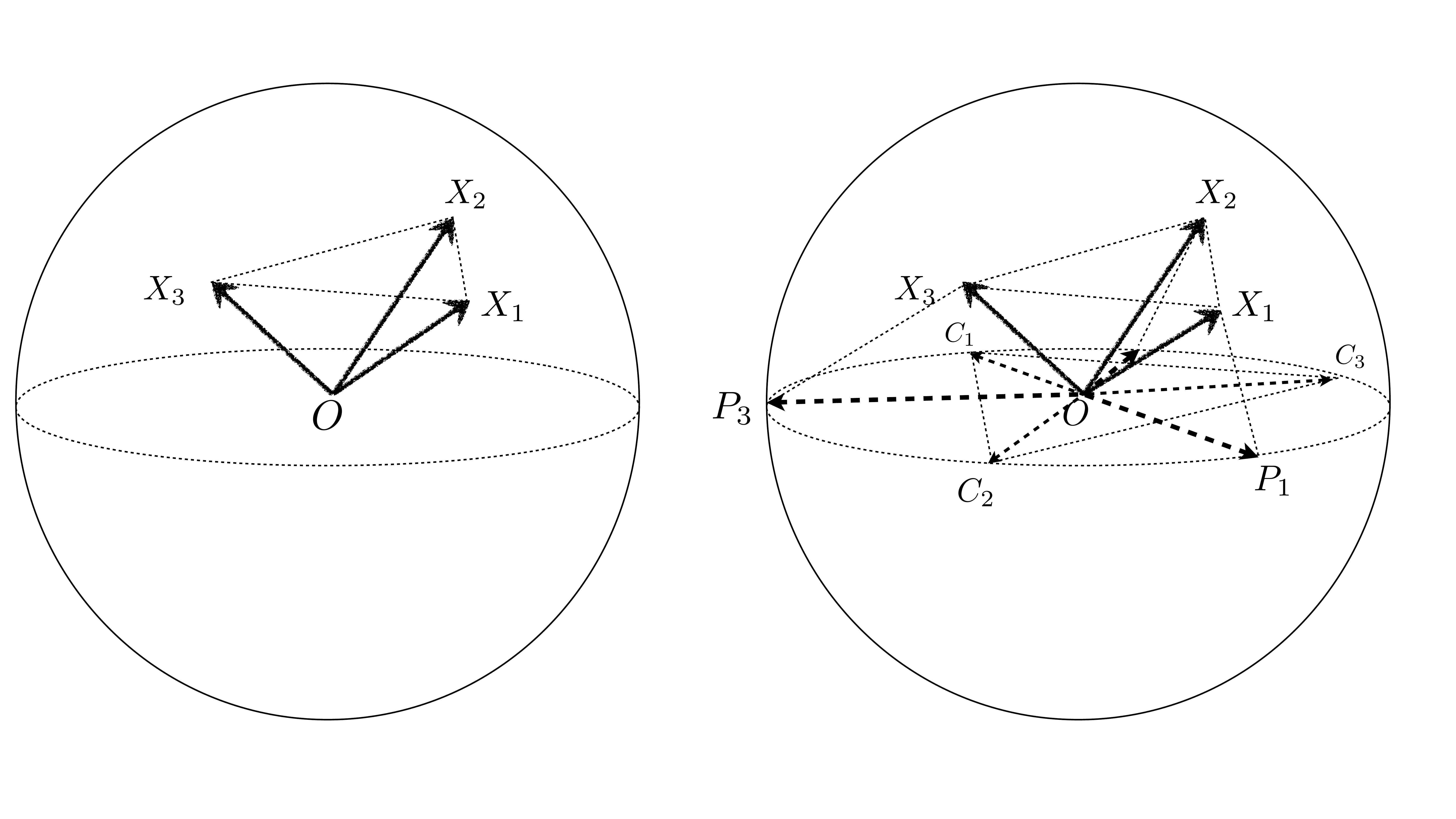}
\caption{ Arbitrary three qubit states are given with equal prior probabilities $1/3$. They are denoted by $OX_{\x}$ for $\x=1,2,3$. The polytope of given states $X_1 X_2 X_3$ defines a plane, and then complementary states and optimal POVM elements lie on the half-plane which is parallel to the defined plane. See the subsection \ref{subsec:three4} for more details. } 
\label{fig:three4}
\end{center}
\end{figure}

Finally, let us show how the geometric formulation in Lemma \ref{lem:4-3} can be generally applied to a set of arbitrary three qubit states when they are given with equal probabilities, $\{1/3,\rho_{x} \}_{\x=1}^{3}$. We explain this, referring to Fig. \ref{fig:three4}. Arbitrary three states are depicted by $OX_{\x}$ for $\x=1,2,3$ respectively.

 Once they are given, they define a plane of the triangle $X_1 X_2 X_3$ in the Bloch sphere, and there exists a half-plane parallel to the plane. To find the guessing probability for the states, one has to first find a maximal triangle that is similar to the triangle $X_1 X_2 X_3$ in the Bloch sphere. Then, the maximal one must lie on the half-plane parallel to $X_1 X_2 X_3$, to fulfill the optimality condition in Eq. (\ref{eq:optpol}), or equivalently Eq. (\ref{eq:symop}). And then, the maximal one is rotated such that the condition in Eq. (\ref{eq:optpol}) is fulfilled, and finally ends up with $C_1 C_2 C_3$. Then, complementary states are immediately found as $OC_{\x}$ for $\x=1,2,3$, from which optimal measurement also follows as those POVM elements orthogonal to complementary states.

\subsection{Four states}
\label{subsec:four}

We now consider discrimination among four qubit states. We begin with a simple case, pairs of orthogonal states. Then, cases where four states define a quadrilateral are considered. Finally, we also show how the geometric formulation can be applied when four states form a tetrahedron in the Bloch sphere.

\subsubsection{Four-state example I: rectangles}

\begin{figure}[t]
\begin{center}
\includegraphics[width=5in]{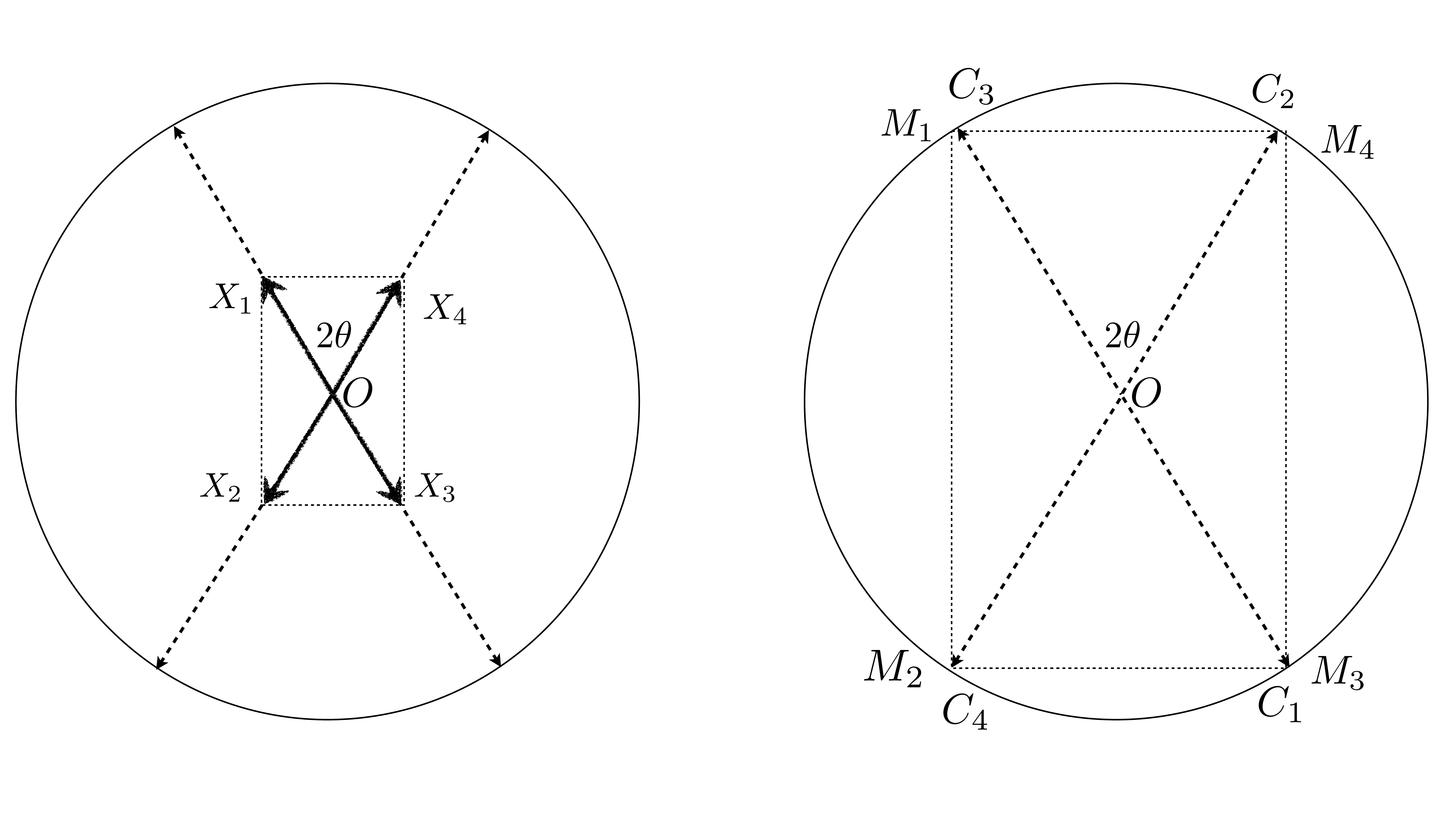}
\caption{ Discrimination of pairs of two orthogonal states, $\{ 1/4, \rho_{\x} = | \psi_{\x} \rangle \langle \psi_{\x}|\}_{\x=1}^{4}$ (see also Eq. (\ref{eq:po})), is shown. Each state $\rho_{\x}/4$ corresponds to $OX_{\x}$ for $\x=1,2,3,4$. Complementary states can be easily found by enlarging and inverting the given polytope $X_1 X_2 X_3 X_4$ so that the optimality condition in Eq. (\ref{eq:r}) is fulfilled. Then, the complementary states form the rectangle $C_1 C_2 C_3 C_4$, from which optimal measurement can also be found as, $\{ M_{\x} \propto OM_{\x}\}_{\x=1}^{4}$, see Eq. (\ref{eq:opt4-1}). The ratio between two rectangles is $1/4$, and thus the guessing probability $1/2$. } 
\label{fig:four1}
\end{center}
\end{figure}

Let us first consider pairs of orthogonal states, say $\{ 1/4,\rho_{\x} = |\psi_{\x} \rangle \langle \psi_{\x}| \}_{\x=1}^{4}$, where $\langle \psi_{1} | \psi_{3} \rangle = \langle \psi_2 | \psi_4 \rangle = 0$. The four states define a plane in the Bloch sphere, and then form a rectangle on the plane. To be explicit, they can be written as follows, for some $\theta_0$ and $\theta$,
\bea
|\psi_1\rangle & = & \cos \frac{\theta_0}{2} |0\rangle + \sin\frac{\theta_0}{2} |1\rangle,  \nonumber \\
|\psi_2\rangle & = & \cos (\frac{\theta_0}{2} - \theta) |0\rangle + \sin (\frac{\theta_0}{2} -\theta) |1\rangle,  \nonumber \\
|\psi_3\rangle & = & \cos (\frac{\theta_0}{2} - \frac{\pi}{2}) |0\rangle + \sin ( \frac{\theta_0}{2} - \frac{\pi}{2} ) |1\rangle,  \nonumber \\
|\psi_4\rangle & = & \cos (\frac{\theta_0}{2} -\theta-  \frac{\pi}{2}) |0\rangle + \sin ( \frac{\theta_0}{2} -\theta- \frac{\pi}{2} ) |1\rangle,  \label{eq:po}
\eea
which are shown in Fig. \ref{fig:four1}. The half-plane in which the four states lie is then defined in the Bloch sphere.

Referring to Fig. \ref{fig:four1}, we apply the geometric formulation in Lemma \ref{lem:4-3}, and show optimal parameters, optimal measurement and complementary states. The polytope of given states is the rectangle $X_1 X_2 X_3 X_4$. It is straightforward to see that the largest one smiler to $X_1 X_2 X_3 X_4$ is the rectangle having its diagonal in length $2$ (in terms of trace-norm). Then, we rotate it so that the relation in Eq. (\ref{eq:optpol}), or the optimality condition in Eq. (\ref{eq:symop}), is satisfied, and then we have the rectangle $C_1 C_2 C_3 C_4$. The complementary states are found as $OC_{\x}$ for $\x=1,2,3,4$, from which optimal POVM elements are also obtained $\{ M_{\x} \propto OM_{\x} \}_{\x=1}^{4}$. Therefore, we have optimal parameters as follows
\bea
&& M_{1} =  \sigma_{3} = | \psi_1 \rangle \langle \psi_1 |, ~~ M_{2} =  \sigma_{4} = | \psi_2 \rangle \langle \psi_2 |, \nonumber\\
&& M_{3}  =  \sigma_{1} = | \psi_3 \rangle \langle \psi_3 |, ~~M_{4} =  \sigma_{2} = | \psi_4 \rangle \langle \psi_4 |. \label{eq:opt4-1}
\eea
From these, the guessing probability can be obtained. Or, applying Lemma \ref{lem:optqdis}, one can see  the ratio, for instance, 
\bea
r = \frac{ X_1 X_2 }{ C_1 C_2} = \frac{OX_1 }{  OC_1} = \frac{1}{4},~\mathrm{and~thus}, ~P_{\g} = \frac{1}{4} + r= \frac{1}{2}.  \label{eq:gfour1}
\eea
We remark that, as it is shown in the above, the guessing probability is equal in the range of angle $0\leq \theta \leq \pi/2$. 

The analysis shown in the above implies that that the guessing probability does not depend on detailed relations among given quantum states but a property assigned by the set of them. This shows the freedom of choosing four states such that the guessing capability about states does not change. More precisely, this can be seen in terms of the symmetry operator, which is in this case,
\bea
K = \frac{1}{4} \rho_{\x}  + r \sigma_{\x} = \frac{1}{2} I \label{eq:K1}
\eea
consistently to Eq. (\ref{eq:gfour1}). That is, all sets of four states of, pairs of orthogonal states, share the same the symmetry operator in Eq. (\ref{eq:K1}), meaning that all of them are in the same equivalence class, see the definition \ref{def:eqc}.

\begin{rem}
\label{rem:po}
Four states of any two pairs of orthogonal qubit states are in the same equivalence class, $\mathcal{A}_{I/2}$.
\end{rem}

\subsubsection{Four-state example II: quadrilateral }

\begin{figure}[t]
\begin{center}
\includegraphics[width=5in]{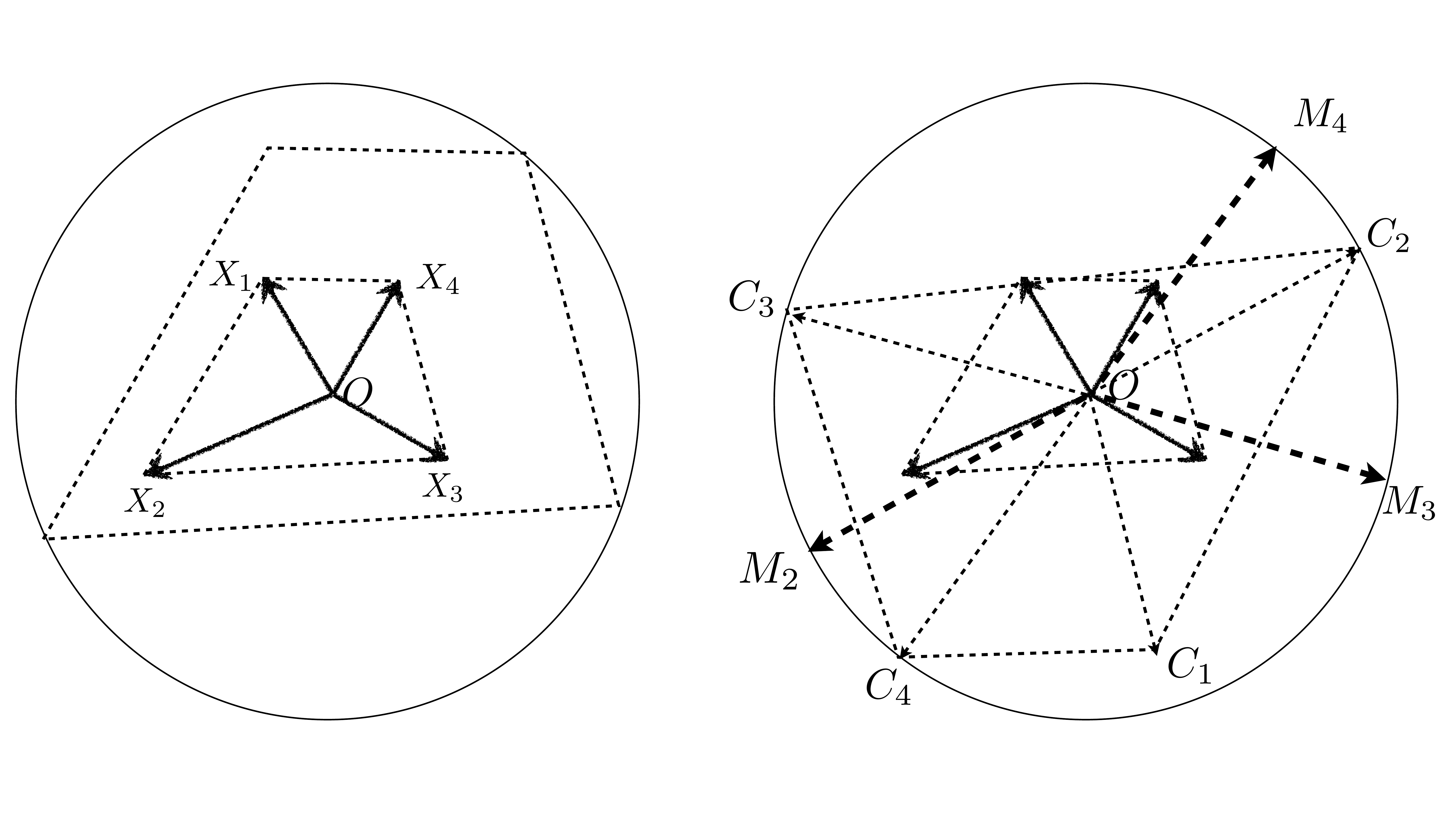}
\caption{ Arbitrary four states lying on a half-plane of the Bloch sphere are given, denoted by $\{q_{\x}=1/4,\rho_{\x} \}_{\x=1}^{4}$. Each state $\rho_{\x}/4$ corresponds to $OX_{\x}$ where $\x=1,2,3,4$. Applying Lemma \ref{lem:4-3}, it is aimed to find a rectangle which is maximal in the Bloch sphere and similar to the given one $X_1X_2X_3X_4$. Then, the maximal rectangle is rotated such that the optimality condition in Eq. (\ref{eq:r}) is satisfied, which then results in $C_1C_2C_3C_4$. Thus, complementary states are found in $OC_{\x}$ for $\x=1,2,3,4$. Note that vertices of $C_1C_2C_3C_4$ may not lie on the surface of the Bloch sphere. For such vertices, optimal POVM elements are the null measurement. Optimal measurement follows straightforwardly as $\{ M_{\x} \propto OM_{\x}\}_{\x=1}^{4}$. } 
\label{fig:four2}
\end{center}
\end{figure}

Next, let us consider arbitrary four qubit states defined on a half-plane, which are given with the equal probabilities $1/4$. For the generality, we do not assume any internal symmetry among given states $\{ 1/4,\rho_{\x}\}_{\x=1}^{4} $ except the assumption that they are on a half-plane in the Bloch sphere. It is then straightforward to generalize this to arbitrary four states defined on any plane in the sphere.

We now refer to Fig. \ref{fig:four2} to show how the geometric formulation in Lemma \ref{lem:4-3} can be applied those four states. The polytope of four states is shown as the quadrilateral $X_1X_2X_3X_4$ where each vertex corresponds to $\rho_{\x}/4$ for $\x=1,2,3,4$. Then, one has to expand the given quadrilateral such that they are maximal in the Bloch sphere. In this way, the ratio between two quadrilaterals can be found.

To find complementary states, one has to rotate the obtained maximal quadrilateral such that the optimality condition in Eq. (\ref{eq:optpol}) is fulfilled. The resulting quadrilateral from which complementary states can be found is then, $C_1C_2 C_3C_4$, where $OC_{\x}$ for $\x=1,2,3,4$ are complementary states. It can happen that, since a given quadrilateral $X_1X_2X_3X_4$ is arbitrarily shaped, some vertices may not be on the surface of the Bloch sphere. As it is depicted in Fig. \ref{fig:four2}, suppose that $OC_1$ cannot lie at the surface of the sphere. This means that complimentary state $\sigma_1$ is not pure, and thus the POVM element for state $\rho_1$ corresponds to the null measurement, $M_{1}=0$. This follows from the optimality condition in Lemma \ref{lem:optqdis} to fulfill Eq. (\ref{eq:orth}). The optimal discrimination strategy for these states is thus to prepare measurement such that there are three kinds of outcomes from $M_{\x}$ with $\x=2,3,4$. Note also that, as the convex hull of the POVM elements, $M_1 M_2M_3$ contains the origin, one can also construct a complement measurement. 


\subsubsection{Four-state example III: tetrahedron }

\begin{figure}[t]
\begin{center}
\includegraphics[width=5in]{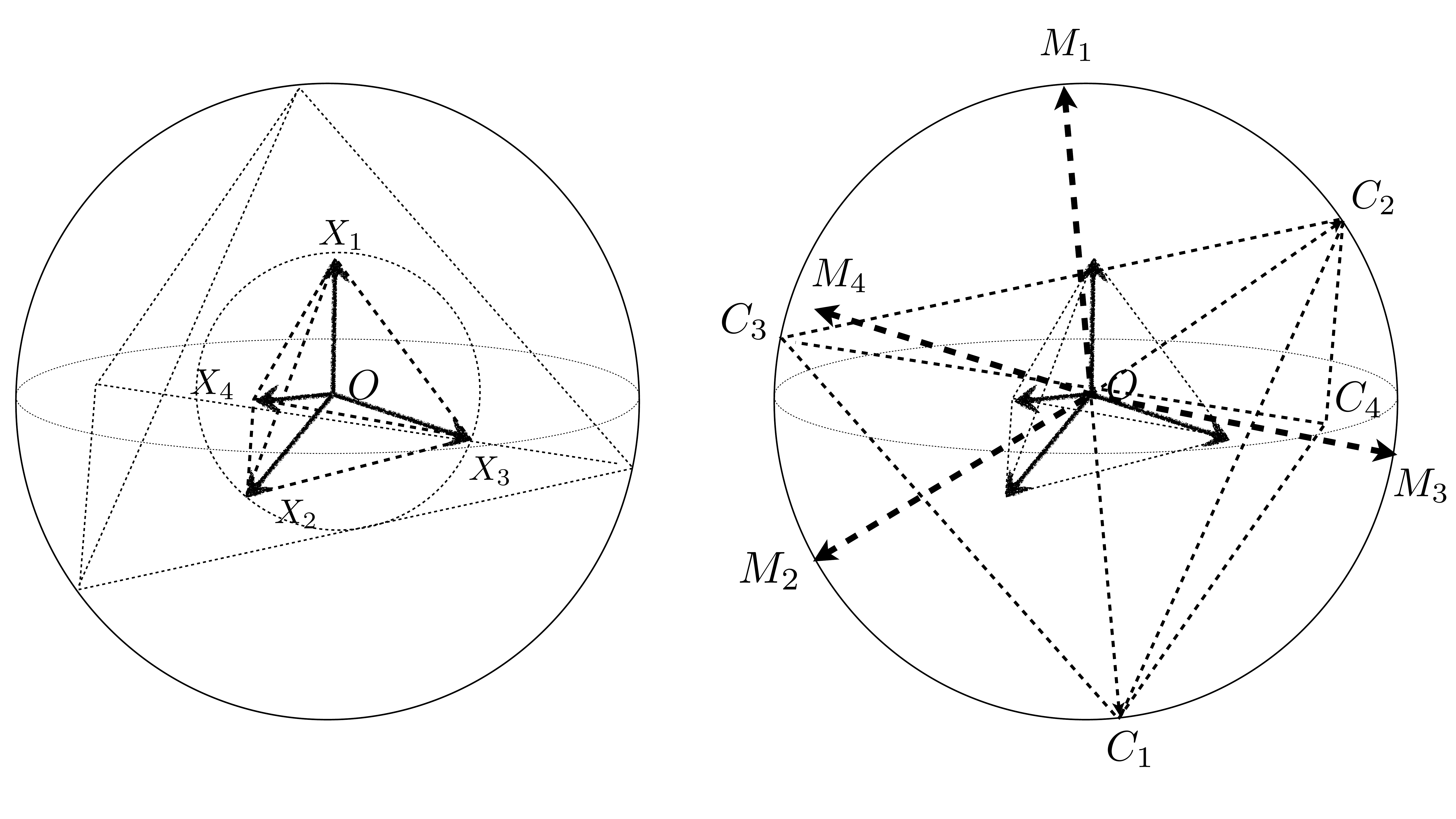}
\caption{It is shown that four given states $OX_{\x}$ ($\x=1,2,3,4$) form a tetrahedron $X_1X_2X_3X_4$ in the Bloch sphere. They are given with prior probabilities $1/4$, and it is not assumed that they are equally spaced one another in the sphere. Applying Lemma \ref{lem:4-3}, one first expands the given tetrahedron until it is maximal in the Bloch sphere, and then rotates the maximal one such that the optimality condition in Eq. (\ref{eq:r}) is fulfilled. The resulting tetrahedron $C_1C_2C_3C_4$ defines the polytope of complementary states, from which it is clear that optimal POVM elements are $\{ M_{\x} = OM_{\x} \}_{\x=1}^{4}$ such that $OM_{\x} = - OC_{\x}$. } 
\label{fig:four3}
\end{center}
\end{figure}

We now consider four qubit states $\{\rho_{\x} \}_{\x=1}^{4}$ which form a polytope having a volume in the Bloch sphere, see Fig. \ref{fig:four3}. Suppose that they are given with probability $1/4$ for each, and for convenience we assume that the tetrahedron constructed by four states is covered by a sphere. Later, this can be relaxed and generalized to cases that all vertices of the tetrahedron are not touched by a sphere. Here, since the tetrahedron is covered by a sphere, four states have the same purity, which we denote by $f$.

Referring to Fig. \ref{fig:four3}, let us show how to find the guessing probability, complementary states and optimal measurement. The guessing probability immediately follows by applying Lemma \ref{lem:4-3}. The ratio $r$ can be computed by finding the ratio of two diameters of two sphere: one is the sphere covering the given tetrahedron, and the other the Bloch sphere covering a maximal tetrahedron similar to the given one. As the purity is given by $f$, the guessing probability is given by
\bea
P_{\g} = \frac{1}{4} + \frac{1}{4}f, \nonumber
\eea
which does not depend on detailed relations among given states such as angles between given states. The above holds true for any four states whose tetrahedron in the Bloch sphere can be covered by a single sphere. 

Then, one rotates the maximal tetrahedron within the Bloch sphere such that the resulting one satisfies the optimality condition in Eq. (\ref{eq:optpol}). Since the tetrahedron already fully occupies the Bloch sphere, it is not difficult to see that the resulting tetrahedron has the reflection symmetry to the given tetrahedron, with respect to the origin of the Bloch sphere, see Fig. (\ref{fig:four3}). Each vertex $OC_{\x}$ for $\x=1,2,3,4$ corresponds to complementary states, and optimal POVM elements are found to be those rank-one operators orthogonal to complementary states

\subsection{Etc.}

So far, we have considered discrimination of two, three, and four qubit states. In this way, for any number of qubit states given, one can apply the geometric formulation and then find the guessing probability, complementary states, and optimal measurement. The framework can be summarized as follows. The first  to do is to construct a polytope of given states, and then one has to search for the polytope of complementary states such that the optimality condition in Eq. (\ref{eq:optpol}), or equivalently Eq. (\ref{eq:symop}), is fulfilled. Two resulting polytopes must be congruent, and correspondingly labeled lines from respective polytopes are anti-parallel. For cases when measurement is applied, its POVM element is of rank-one and also corresponding complementary states must be of rank-one, that is, pure states. If it is found that complementary states are not of rank-one, then their POVM elements are the null operator and thus the optimal strategy for those states is to apply no-measurement. \emph{For the implementation in experiment, this means that no output port is needed for those states}. 

As we have also shown, there is a further simplification in optimal discrimination when the polytope of quantum states given with equal prior probabilities is tightly covered by a sphere, such that all vertices of the polytope touch the sphere. Then, the guessing probability simply follows from the ratio between diameters of two spheres, one from the sphere covering a given polytope and the other the Bloch sphere. In this case, it is explicitly shown that the guessing probability does not depend on detailed relations of given states such as angles between any two given states. 

\begin{rem}
For $N$ qubit states given with equal prior probabilities $1/N$, if their polytope in the Bloch sphere is covered by a smaller sphere of radius $r$ such that each vertex touches the sphere and contains the origin of the Bloch sphere, the guessing probability for those states is given by $P_{\g} = 1/N + r$ independently to detailed relations among given states.
\end{rem}

For high-dimensional quantum systems, a general geometric expression is lacked and therefore the geometric formation shown in the subsection \ref{subsec:geofom} cannot be further applied in general. Once given states give an underlying geometry in which the convex polytope picture is clear, one can apply and formalize the method of the geometric formulation.

\section{Conclusion}
\label{sec:con}

In this work, we have considered the problem of ME state discrimination and shown the general structure of the problem. The main idea of the development is to view the problem from various approaches. Finally, it turns out hat the general structure can be summarized in the formulation of the so-called complementarity problem that generalizes convex optimization. The key element in the structure is a single positive operator, called as the symmetry operator, which gives the complete characterization of optimal discrimination. Then, one can exploit the symmetry operator to find optimal parameters in the ME discrimination such as the guessing probability, complementary states and optimal measurement. The symmetry operator also allows an interpretation to the guessing probability as the averaged distance of given states being deviated from the symmetry operator in terms of the trace norm. The interpretation is in accordance to cases when classical systems are employed, where the guessing probability is interpreted as the deviation from the uniform distribution. 

It is shown that in ME discrimination of quantum states, the symmetry operator is uniquely determined whereas optimal measurement is not. This means, rather than optimal measurement, a symmetry operator can characterize the ME discrimination. Symmetry operators are therefore exploited to define equivalence classes among sets of quantum states, such that for those sets in the same class, the ME discrimination is completely characterized by an identical symmetry operator. This provides an alternative approach to ME discrimination: by checking whether given two sets are in the same class, one can find optimal discrimination. In the other way around, given a symmetry operator, we have shown how one can construct a set of quantum states for which the ME discrimination is characterized by the operator. 

From general structures found from the optimality conditions, we have provided a geometric formulation of ME state discrimination. In the formulation, geometry of quantum states is exploited to find the guessing probability, instead of optimization over measurement. More precisely, the polytope of given states in the state space is linked to the guessing probability, without directly referring to measurement operators via the measurement postulate. It is clear that the method can be applied once the underlying geometry of given states is well-defined. Conversely, we have also argued that, from cases where optimal discrimination is known, the guessing probability is useful to find the underlying geometry of high-dimensional quantum states. We have applied the geometric formulation to qubit states and solve ME discrimination: i) the complete solution is provided for any set of qubit states when prior probabilities are equal, ii) this gives an upper bound to cases when prior probabilities are not equal, iii) solutions are obtained even if given states do not contain any symmetry among them, iv) it is shown that the guessing probability does not depend on detailed relations among given states but a geometric property assigned by the set itself. The conclusion iv) is along the conclusion in Ref. \cite{ref:jozsa} that distinguishability within an ensemble of quantum states is assigned as a global property that cannot be reduced to properties of pairs of states. We here arrive at the conclusion by quantifying the distinguishability with the guessing probability, while it is with von Neumann entropy in Ref. \cite{ref:jozsa}.

Discrimination of quantum states poses a simple question connected to fundamentals and profound limitations in various contexts of quantum information theory. The results presented here not only provide a useful method of solving optimal discrimination, but also give a general, unique, and fresh understanding to ME quantum state discrimination. 


\section*{Acknowledgement}

The author thanks J. Bergou, B.-G. Englert, T. Fritz, D. Gross, W.-Y. Hwang, L.-C. Kwek, A. Monras, M. Navascu\'es, H. K. Ng, P. Raynal, S. Yun, S. Wehner, and A. Winter, H. Zhu for helpful discussions and comments while preparing the manuscript. This work is supported by National Research Foundation and Ministry of Education, Singapore.



\begin{thebibliography}{99}

\bibitem{ref:holevo} A. S. Holevo, Probl. Inf. Transf. \textbf{10}, 317 (1974).

\bibitem{ref:yuen} H. P. Yuen, R. S. Kennedy, and M. Lax, IEEE Trans. Inf. Theory \textbf{21}, 125 (1975).

\bibitem{ref:hel} C. Helstrom, Quantum Detection and Estimation Theory Academic, New York, ADDRESS, (1976).

\bibitem{ref:sas} M. Sasaki, S. M. Barnett, R. Jozsa, M. Osaki and O. Hirota: Phys. Rev. A \textbf{59}, 3325 (1999).

\bibitem{ref:iva} D. Ivanovic, Phys. Lett. A \textbf{123}, 257 (1987).

\bibitem{ref:die} D. Dieks, Phys. Lett. A \textbf{126}, 303 (1988).

\bibitem{ref:per} A. Peres, Phys. Lett. A \textbf{128}, 19 (1988).

\bibitem{ref:jae} G. Jaeger and A. Shimony, Phys. Lett. A \textbf{197}, 83 (1995).

\bibitem{ref:ban} M. Ban, Phys. Lett. A \textbf{213}, 235 (1996).

\bibitem{ref:cro} S. Croke, E. Andersson, S. M. Barnett, C. R. Gilson, J. Jeffers, Phys. Rev. Lett. \textbf{96}, 070401 (2006). 

\bibitem{ref:bagan} E. Bagan, R. Munoz-Tapia, G. A. Olivares-Renteria, J. A. Bergou, arXiv:1206.4145.

\bibitem{ref:herzog} U. Herzog, arXiv:1206.4412. 

\bibitem{ref:rev1} A. Chefles, Contemporary Physics \textbf{41}, 401 (2000).

\bibitem{ref:rev2} J. A. Bergou, U. Herzog, and M. Hillery, Lect. Notes Phys. 649, 417-465 (Springer, Berlin, 2004).

\bibitem{ref:rev3} J. A. Bergou, J. Phys: Conf. Ser. \textbf{84}, 012001 (2007). 

\bibitem{ref:rev4} S. M. Barnett and S. Croke, Adv. Opt. Photon. \textbf{1}, 238Ð278 (2009).

\bibitem{ref:rev5} J. A. Bergou, Journal of Modern Optics, Vol. \textbf{57}, Issue 3, 160-180 (2010).
















\bibitem{ref:jez} M. Jezek, J. Rehacek, and J. Fiurasek, Phys. Rev. A \textbf{65}, 060301(R) (2002).

\bibitem{ref:eldarsdp2} Y. C. Eldar, IEEE Trans. Inf. Theory {\bf 49}, 446Ð456 (2003).

\bibitem{ref:eldar-sdp} Y. C. Eldar, A. Megretski, and G. C. Verghese, IEEE Trans. Inf. Theory \textbf{50}, 1198 (2004).

\bibitem{ref:sdpbook} S. Boyd and L. Vandenberghe, Convex Optimization, Cambridge University Press, (2004).

\bibitem{ref:mirror3} A. Chefles and S. M. Barnett, Phys. Lett. A {\bf 250}, 223Ð229 (1998).

\bibitem{ref:eldar} Y. C. Eldar and G. D. Forney, IEEE Trans. Inf. Theory \textbf{47}, 3 (2001).

\bibitem {ref:mirror0} S. M. Barnett, Phys. Rev. A {\bf 64}, 030303 (2001).

\bibitem{ref:mirror4} J. Mizuno, M. Fujiwara, M. Akiba, T. Kawanishi, S. M. Barnett and M. Sasaki, Phys. Rev. A {\bf 65}, 012315 (2001).

\bibitem{ref:mirror1} E. Andersson, S. M. Barnett, C. R. Gilson and K. Hunter, Phys. Rev. A {\bf 65}, 052308 (2002). 

\bibitem{ref:mirror2} C.-L. Chou, Phys. Rev. A {\bf 70}, 062316 (2004).



\bibitem{ref:com} K. G. Murty, Linear complementarity, linear and nonlinear programming, Heldermann Verlag (1998). 

\bibitem{ref:maurer} S. Dziembowski and U. Maurer, Journal of Cryptology, vol. 17, no. 1, pp. 5Ð26 (2004).


\bibitem{ref:bae} J. Bae and W.-Y. Hwang, Phys. Rev. A \textbf{87}, 012334 (2013).

\bibitem{ref:jozsa} R. Jozsa and J. Schlienz, Phys. Rev. A \textbf{62}, 012301 (2000).

\bibitem{ref:bae11} J. Bae, W.-Y. Hwang, and Y.-D. Han, Phys. Rev. Lett. \textbf{107}, 170403 (2011).

\bibitem{ref:schr} E. Schr\"odinger, Proc. Camb. Phil. Soc. \textbf{31}, 553 (1935), \emph{ibid}. \textbf{32}, 446, (1936).

\bibitem{ref:ghjw1} N. Gisin, Helv. Phys. Acta \textbf{62}, 363 (1989).

\bibitem{ref:ghjw2} L. P. Hughston, R. Jozsa, and W. K. Wootters, Phys. Lett. A \textbf{183}, 14 (1993).

\bibitem{ref:mochon} For a specific example, one can refer to the case of linearly dependent quantum states, e.g. Mochon, Phys. Rev. A \textbf{73}, 032328 (2006).

\bibitem{ref:eldarvn} Y. C. Eldar, Phys. Rev. A \textbf{68}, 052303 (2003).

\bibitem{ref:terhal} M. E. Deconinck and B.M. Terhal, Phys. Rev. A \textbf{81}, 062304 (2010).

\bibitem{ref:hunter} K. Hunter, Phys. Rev. A \textbf{68}, 012306 (2003).





\end{thebibliography}
\end{document}